\documentclass[twocolumn,showpacs,amsmath,amssymb,superscriptaddress]{revtex4-1}
\usepackage{dcolumn}
\usepackage{color}
\usepackage{graphicx}
\usepackage{amsmath}
\usepackage{multirow}
\usepackage{bm}
\usepackage{url}

\newcommand\+{\dagger}

\begin{document}

\title{Prolate-to-oblate shape phase transitions in neutron-rich odd-mass nuclei}

\author{K.~Nomura}
\affiliation{Physics Department, Faculty of Science, University of
Zagreb, HR-10000 Zagreb, Croatia}

\author{R.~Rodr\'iguez-Guzm\'an}
\affiliation{Physics Department, Kuwait University, 13060 Kuwait, Kuwait}

\author{L.~M.~Robledo}
\affiliation{Departamento de F\'\i sica Te\'orica, Universidad
Aut\'onoma de Madrid, E-28049 Madrid, Spain}

\affiliation{Center for Computational Simulation,
Universidad Polit\'ecnica de Madrid,
Campus de Montegancedo, Boadilla del Monte, 28660-Madrid. Spain
}

\date{\today}

\begin{abstract}	
We investigate the prolate-to-oblate shape phase transitions in the 
neutron-rich Pt, Os and Ir nuclei in the mass $A\approx 190$ region. 
The Hamiltonian of the interacting boson-fermion model, used to describe 
the odd-mass $^{185-199}$Pt, $^{185-193}$Os and $^{185-195}$Ir 
isotopes, is partially constructed by using as a microscopic input the results of 
constrained self-consistent mean-field calculations within the 
Hartree-Fock-Bogoliubov method with the Gogny force. The remaining few 
parameters are adjusted to experimental data in the odd systems. In this
way the calculations reasonably describe 
the spectroscopic properties of the odd-mass systems considered. 
Several calculated observables for the odd-mass nuclei, especially the 
low-energy excitation spectra and the effective deformation parameters, 
point to a prolate-oblate shape transition as a function of the 
neutron number for all the isotopic chains considered and similar to the one 
already observed in the neighboring even-even systems.
\end{abstract}


\maketitle


\section{Introduction}


One of the most prominent features of the atomic nucleus is that it 
organizes itself into various types of geometrical shapes \cite{BM} that often 
evolve gradually as  functions of the nucleon number within an isotopic 
or isotonic chain. In some instances, however, such  shape evolution 
takes place abruptly at particular nucleon numbers. This phenomenon is 
known as a (quantum) shape phase transition \cite{cejnar2010}. Over the 
past decades, numerous experiments have been carried out to measure  
observables signaling such  phase transitions \cite{cejnar2010}. 
Theoretical calculations have also been carried out within several  
frameworks \cite{niksic2007,robledo2009,li2010,cejnar2010,shimizu2017}. 
Typical examples of shape phase transitions are, the 
spherical-to-axially-deformed \cite{iachello2001} and the 
spherical-to-$\gamma$-soft \cite{iachello2000} ones. Other types of  
transitions include the one that occurs between prolate and oblate 
configurations going through a transitional $\gamma$-soft shape 
\cite{jolie2001}.

Nuclear shape transitions have been  well studied for even-even nuclei. 
There is, however, a wealth of experimental information for odd-mass 
systems that remains to be analyzed from a theoretical perspective. 
Within this context, it is particularly interesting to 
consider the nature of  phase transitions in those odd-mass nuclei and 
how they correlate with the ones in the neighbouring even-even systems 
\cite{iachello2011}. However, the theoretical description of  odd-mass 
nuclei tends to be more cumbersome than for even-even systems, as one 
needs to take into account both collective and single-particle degrees 
of freedom on an equal footing \cite{bohr1953}.

The aim of this paper is to study the effect of the odd particle on the 
prolate-to-oblate shape transition in  neutron-rich nuclei with mass 
number $A\approx 190$. To this end, we have selected the  odd-mass  
systems $^{185-199}$Pt $^{185-193}$Os and $^{185-195}$Ir. Their 
even-even neighbors $^{186-200}$Pt and $^{186-194}$Os, are considered 
to be good examples of the prolate-to-oblate shape transition. In many cases 
$\gamma$-soft shapes  are also found. Therefore, they represent  a 
stringent test for nuclear structure models. In order to describe  
spectroscopic properties, we have  resorted to the recently developed 
method of Ref.~\cite{nomura2016odd}, based on the nuclear energy 
density functional  (EDF) theory and the particle-core coupling scheme 
\cite{iachello1979,IBFM}. The method has already been applied to study the 
spherical-to-axially-deformed \cite{nomura2016qpt,nomura2017odd-2} and 
spherical-to-$\gamma$-soft \cite{nomura2017odd-1,nomura2017odd-3} shape 
phase transitions as well as octupole correlations in neutron-rich Ba 
nuclei \cite{nomura2018oct}. The robustness of the method has been 
studied using both non-relativistic \cite{nomura2017odd-2} and 
relativistic \cite{nomura2017odd-1} EDFs.

In this work, the even-even Pt and Os nuclei are described within the 
neutron-proton interacting boson model (IBM-2) \cite{OAI,IBM} built on 
the neutron (proton) $s_{\nu}$ and $d_{\nu}$ ($s_{\pi}$ and $d_{\pi}$) 
bosons, which represent correlated $J^{\pi}=0^+$ and $2^+$ pairs of 
valence neutrons (protons) \cite{OAI}. On the other hand, the 
particle-core coupling is considered within the neutron-proton 
interacting boson-fermion model (IBFM-2) \cite{alonso1984,IBFM}. Similar to our previous 
studies \cite{nomura2017odd-2,nomura2017odd-3}, which were based on the 
simpler IBM-1 model \cite{nomura2017odd-2}, the strength parameters for 
the IBM-2 Hamiltonian, the single-particle energies and  the occupation 
probabilities of the odd particle, are determined by  constrained 
self-consistent mean-field calculations based on the Gogny-D1M  EDF 
\cite{D1M,Gogny}. The coupling constants of the boson-fermion 
interaction are the only free parameters of the model. They are specifically fitted 
to  reproduce the low-lying excitation spectrum for each odd-mass 
nucleus. The IBFM-2 phenomenology has already been considered  in this 
region of the nuclear chart \cite{arias1986}. However, in this work we 
resort to a microscopic input obtained from the Gogny-D1M EDF 
framework, i.e., the IBM-2 Hamiltonian parameters, single-particle 
energies and occupation probabilities are  determined within the HFB 
framework. In a previous study \cite{nomura2011sys}, we have already 
considered even-even nuclei in this mass region, including the Pt and 
Os ones studied here, with the IBM-2 Hamiltonian parameters derived 
from HFB calculations based on the  Gogny-D1M EDF. In this work, we 
will take the IBM-2 Hamiltonian parameters for  the even-even nuclei 
from Ref~\cite{nomura2011sys} and focus on the remaining ones to study the 
odd-mass nuclei. Let us also mention that other theoretical frameworks, 
like the symmetry-projected generator coordinate method (GCM) for odd 
mass systems \cite{bally2014,borrajo2016} and the large-scale shell 
model \cite{caurier2005,shimizu2017}, could be employed. However, they 
are computationally much more demanding for heavier and/or open-shell 
nuclei. Hence, computationally feasible schemes, such as the 
particle-vibration coupling \cite{bohr1953}, represent a more feasible 
alternative and have often been considered in the literature, e.g., \cite{colo2017}.

The paper is organized as follows. In Sec.~\ref{sec:model}, we briefly 
outline the theoretical framework used in this study. There, we will 
also discuss the Gogny-HFB deformation energy surfaces as well as  the  
parameters of the Hamiltonian. Then, in Sec.~\ref{sec:sys}, we discuss  
the spectroscopic properties of the considered nuclei. We briefly 
review the results obtained for even-even nuclei in 
Sec.~\ref{sec:even}. The systematics of the low-lying yrast levels in 
the odd-mass nuclei is presented in Sec.~\ref{sec:odd}. More detailed 
level schemes and electromagnetic properties for  some selected 
odd-mass nuclei are discussed in Sec.~\ref{sec:detail}. As yet another 
signature of the prolate-to-oblate shape transition, in 
Sec.~\ref{sec:def}, we consider effective $\beta$ and $\gamma$ 
deformations. Finally, Sec.~\ref{sec:summary} is devoted to the 
concluding remarks.


\section{Building the interacting boson-fermion Hamiltonian\label{sec:model}}


The IBFM-2 Hamiltonian is comprised of the IBM-2 Hamiltonian 
$\hat H_{\rm B}$ \cite{nomura2011sys}, the Hamiltonian for the odd nucleon $\hat H_{\rm F}$,
and the boson-fermion interaction $\hat H_{\rm BF}$: 
\begin{eqnarray}
\label{eq:ham}
 \hat H_{\rm IBFM} = \hat H_{\rm B} + \hat H_{\rm F} + \hat H_{\rm BF}.
\end{eqnarray}

In this expression, the doubly-magic nucleus $^{208}$Pb is taken as the 
inert core. In the IBM-2, the number of neutron (proton) bosons 
$N_{\nu}$ ($N_\pi$) is equal to half the number of valence neutrons 
(protons) and is counted as the number of holes in the latter half of a 
given major shell. In the present case, all the bosons are  hole-like 
and therefore $2\leq N_\nu\leq 9$ and $N_\pi= 2$ for $^{186-200}$Pt and 
$4\leq N_\nu\leq 8$ and $N_\pi= 3$ for $^{186-194}$Os. The strength 
parameters for the IBM-2 Hamiltonian for the even-even nuclei 
$^{186-200}$Pt and $^{186-194}$Os have been previously determined 
\cite{nomura2008} by mapping  the $(\beta,\gamma)$-deformation energy 
surface, computed within the constrained Gogny-D1M HFB approach, onto 
the expectation value of the IBM-2 Hamiltonian in the boson condensate 
state \cite{ginocchio1980}. For a more detailed account of the whole 
procedure, the reader is referred to 
Refs.~\cite{nomura2008,nomura2011pt,nomura2011sys}. The parameters 
obtained via the mapping procedure can be found in Table I of 
Ref.~\cite{nomura2011sys}.


\begin{figure}[htb!]
\begin{center}
\includegraphics[width=\linewidth]{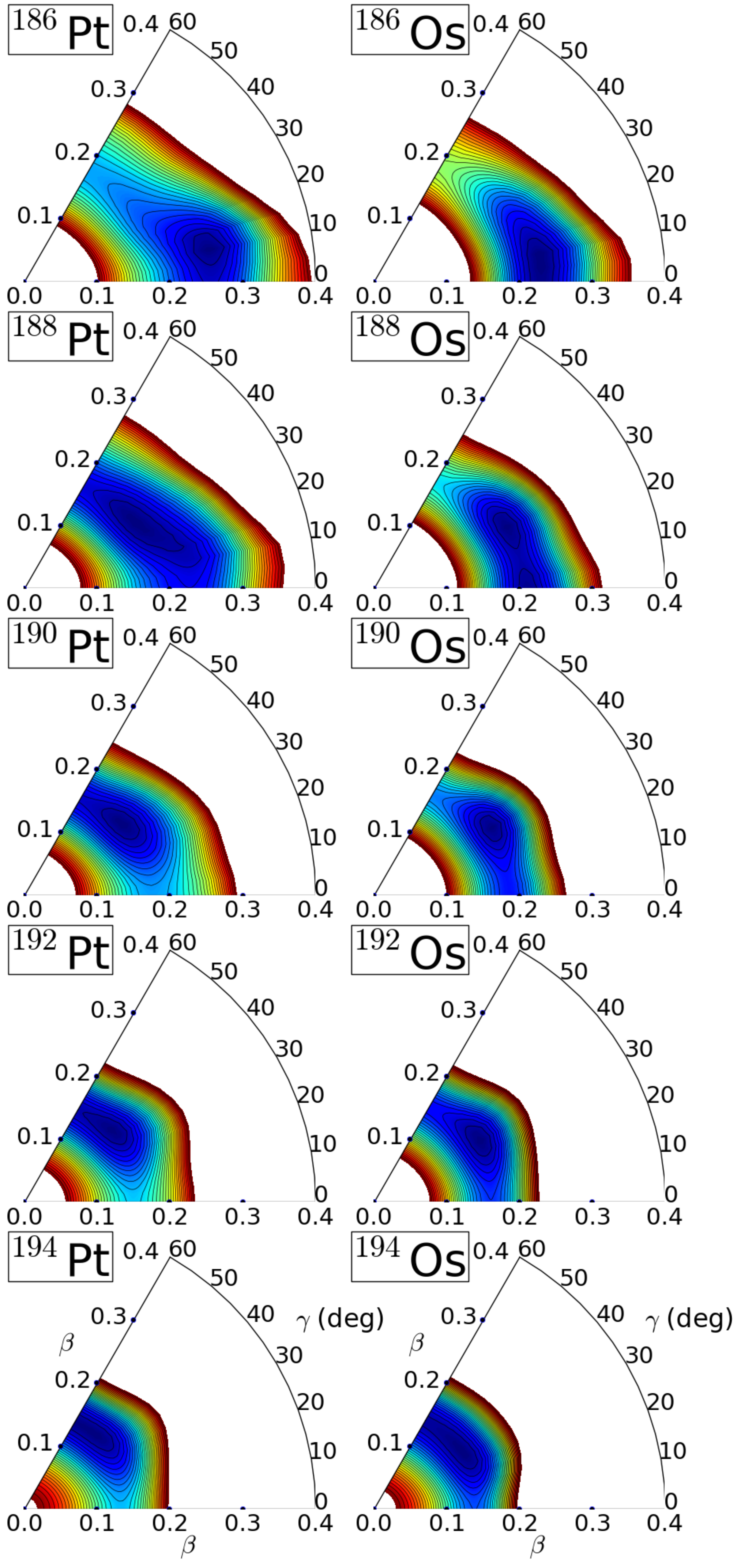}
\caption{(Color online) The Gogny-D1M HFB deformation energy surfaces in the
 $(\beta,\gamma)$-deformation space for the 
 $^{186-194}$Pt and  $^{186-194}$Os nuclei are plotted  up to 3 MeV from the
 global minimum. The energy difference between the neighbouring
 contours is 100 keV.  
 } 
\label{fig:pes}
\end{center}
\end{figure}

In Fig.~\ref{fig:pes} we have depicted the Gogny-D1M energy surfaces 
for those even-even Pt and Os nuclei corresponding to the 
prolate-oblate transitional regions. Results for other nuclei can be 
found in Ref.~\cite{robledo2009}. Note, that in Ref.~\cite{robledo2009} 
we have resorted to the Gogny-D1S \cite{D1S} EDF. However, the 
mean-field surfaces obtained with the parametrization D1S are 
essentially the same as the ones provided by the  parameter set D1M. 
The minimum of each of the energy surfaces in the figures changes 
gradually, as a function of neutron number, from near prolate to 
shallow triaxial  (around $^{188}$Pt and $^{190}$Os) and then near 
oblate  (around $^{192}$Pt and $^{194}$Os). The IBM-2 energy surfaces, 
obtained via the mapping procedure, can be found in 
Ref.~\cite{nomura2011sys}.

The second term in Eq.~(\ref{eq:ham}) reads, $\hat H_{\rm 
F}=\sum_{j}\epsilon_{j\tau} (a_{\tau,j}^{\dagger}\times \tilde 
a_{\tau,j})^{(0)}$, with $\epsilon_j$ being the single-particle 
energies of the orbitals for the odd neutron ($\tau=\nu$) or proton 
($\tau=\pi$). As fermionic valence space for the odd-$N$ Pt and Os 
nuclei, we have taken the whole neutron $N=82-126$ major shell: 
$3p_{1/2}$, $3p_{3/2}$, $2f_{5/2}$, $2f_{7/2}$, $1h_{9/2}$ for 
negative-parity states and $1i_{13/2}$ for positive-parity states. For 
the odd-$Z$ Ir isotopes we have taken the whole proton $Z=50-82$ major 
shell: $3s_{1/2}$, $2d_{3/2}$, $2d_{5/2}$, $1g_{7/2}$ for positive 
parity and $1h_{11/2}$ for negative-parity states.  Note that all the 
valence particles are treated here as holes. Therefore, for an odd-mass 
nucleus with mass $A$, its even-even neighbor with mass $A+1$ is taken 
as the even-even boson core.


\begin{figure}[htb!]
\begin{center}
\includegraphics[width=\linewidth]{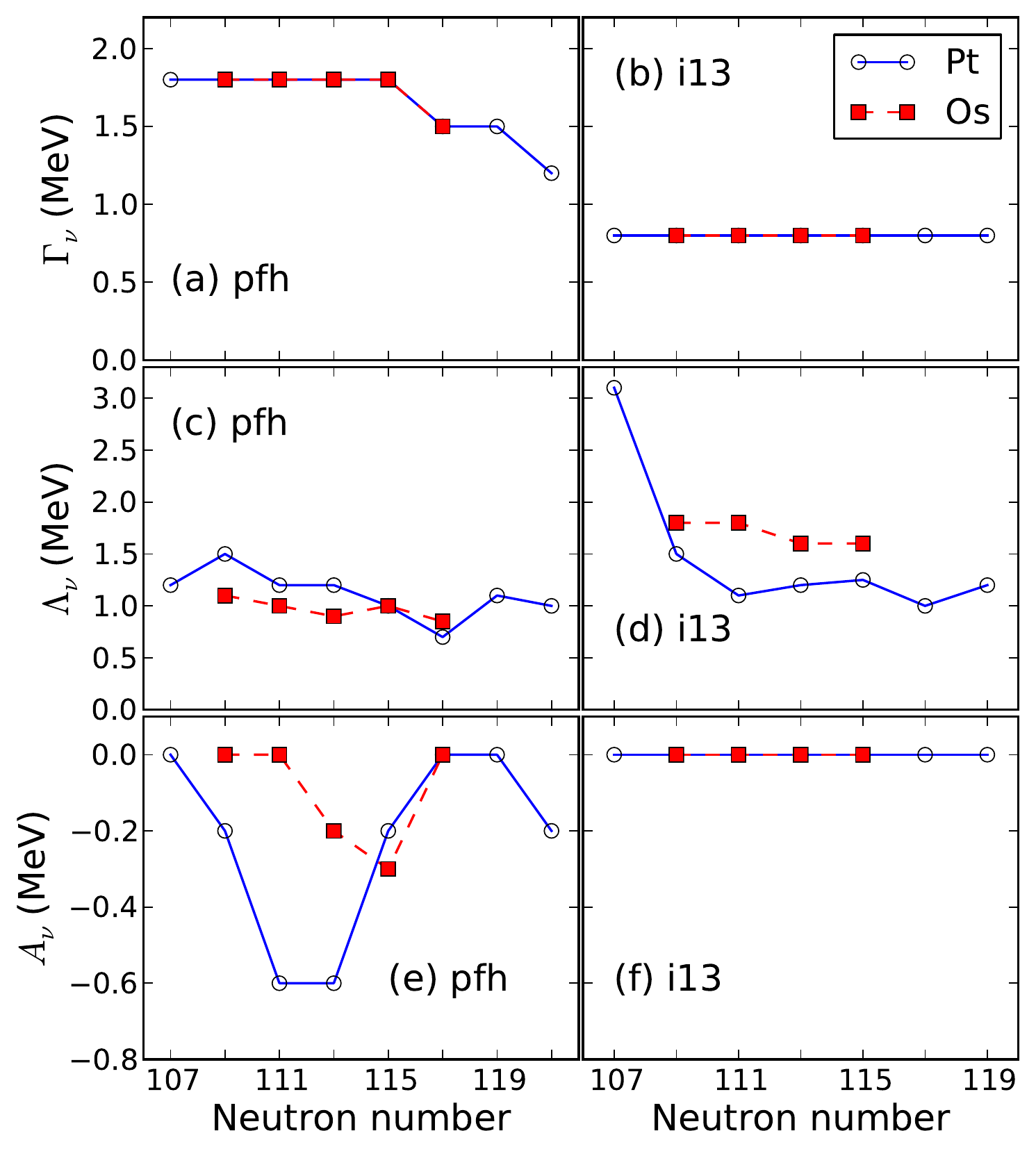}
\caption{(Color online) The strength parameters
 $\Gamma_\nu$, $\Lambda_\nu$ and $A_\nu$, employed
 for the odd-$N$ Pt and Os nuclei are depicted  as functions of the neutron number.} 
\label{fig:para-ptos}
\end{center}
\end{figure}


\begin{figure}[htb!]
\begin{center}
\includegraphics[width=\linewidth]{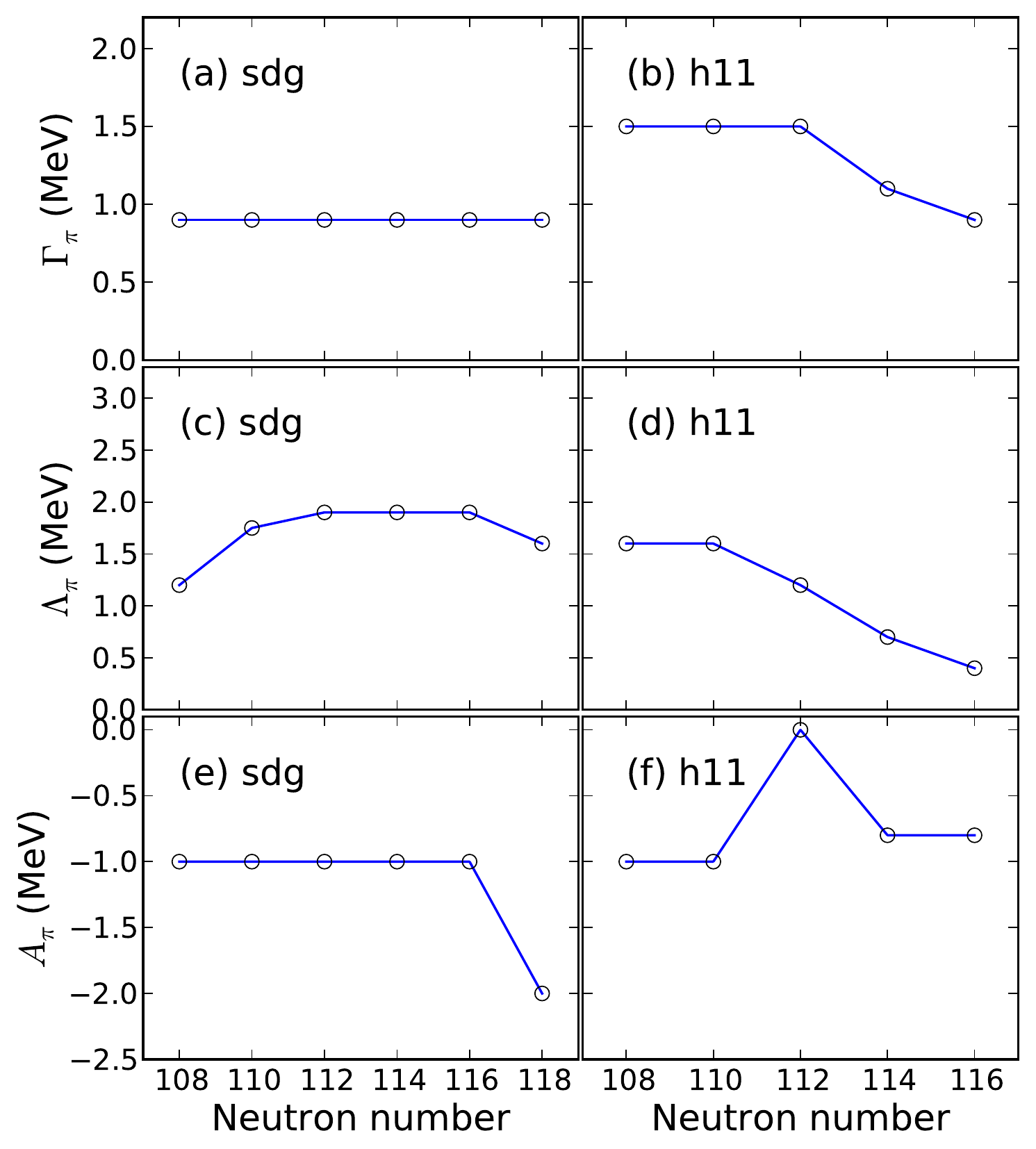}
\caption{(Color online) The strength parameters $\Gamma_\pi$,
 $\Lambda_\pi$ and $A_\pi$ for the
 odd-$Z$ Ir isotopes are depicted  as functions of the neutron number. } 
\label{fig:para-ir}
\end{center}
\end{figure}


\begin{figure}[htb!]
\begin{center}
\includegraphics[width=\linewidth]{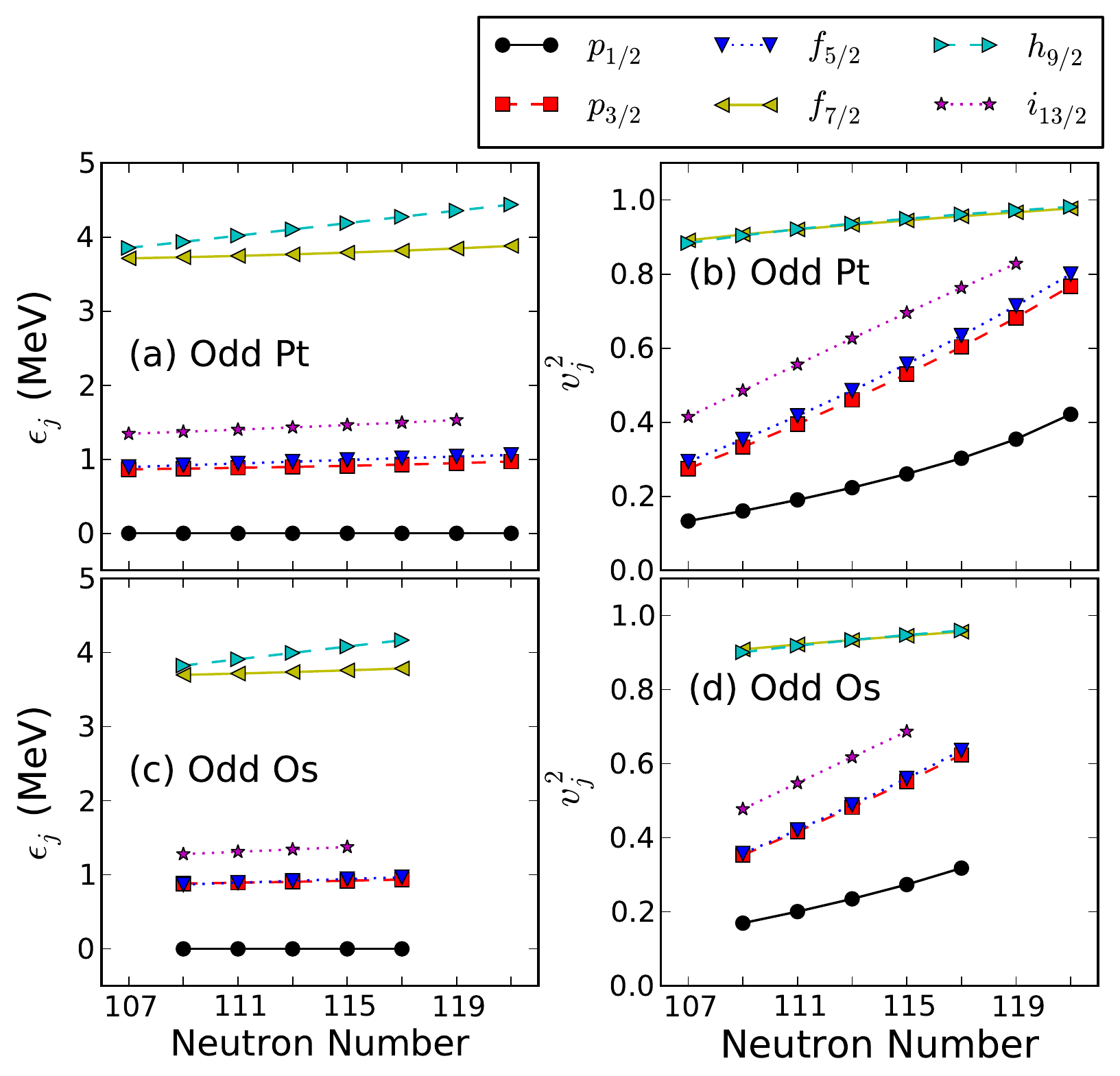}
\caption{(Color online) Single-particle energies (plotted with respect to
 that of the $3p_{1/2}$ orbital) and occupation
 probabilities employed for the odd-$N$ Pt and Os nuclei as functions of the neutron number. } 
\label{fig:spe-ptos}
\end{center}
\end{figure}


\begin{figure}[htb!]
\begin{center}
\includegraphics[width=\linewidth]{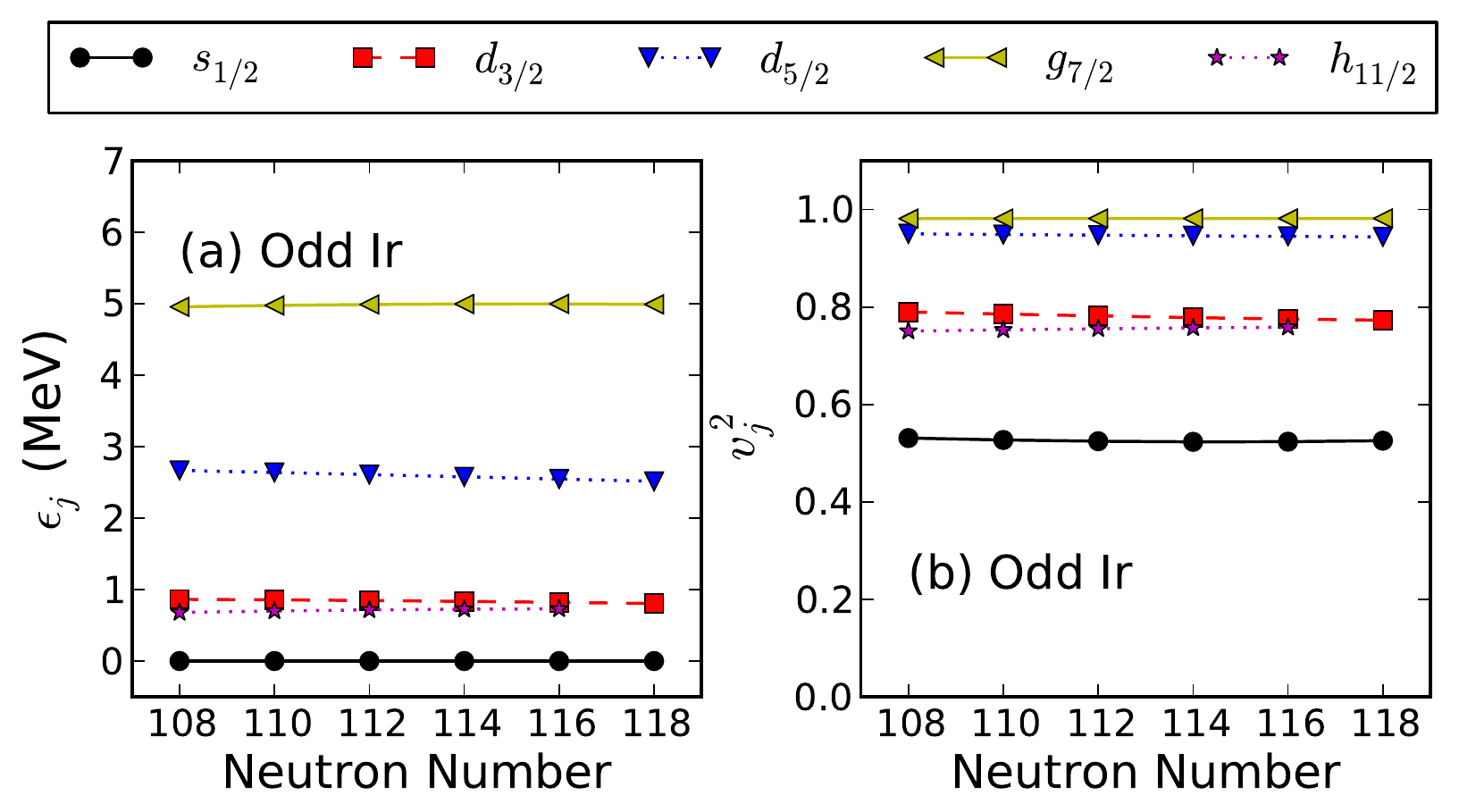}
\caption{(Color online) The same as in  Fig.~\ref{fig:spe-ptos} but for the
 odd-$Z$ Ir isotopes. The single-particle energies are plotted with respect to
 $3s_{1/2}$ orbital.} 
\label{fig:spe-ir}
\end{center}
\end{figure}

For the boson-fermion interaction term $\hat H_{\rm BF}$ in
Eq.~(\ref{eq:ham}), we use an expression similar to the one used in previous studies
\cite{scholten1985,arias1986}: 
\begin{eqnarray}
\label{eq:ham-bf}
 \hat H_{\rm BF} = &&\Gamma_\nu\hat Q_{\pi}^{(2)}\cdot\hat q_{\nu}^{(2)} +
 \Gamma_\pi\hat Q_{\nu}^{(2)}\cdot\hat q_{\pi}^{(2)}  \nonumber \\ 
&&+
  \Lambda_\nu\hat V_{\pi\nu} + \Lambda_\pi\hat V_{\nu\pi} + A_\nu\hat n_{d\nu}\hat
  n_{\nu} + A_\pi\hat n_{d\pi}\hat
  n_{\pi}, 
\end{eqnarray}
where  the first and second terms are the quadrupole dynamical terms,
with the bosonic 
quadrupole operator for proton $\hat Q^{(2)}_{\pi}$  and 
neutron $\hat Q^{(2)}_{\nu}$, respectively. The fermionic quadrupole 
operator for the odd neutron or proton reads: 
\begin{eqnarray}
\hat
q^{(2)}_\tau=\sum_{jj'}\gamma_{jj'}(a^\+_{j\tau}\times\tilde
a_{j'\tau})^{(2)},
\end{eqnarray} 
where $\gamma_{jj'}=(u_ju_{j'}-v_jv_{j'})Q_{jj'}$ and  $Q_{jj'}=\langle
j||Y^{(2)}||j'\rangle$ represents
the matrix element of the fermionic 
quadrupole operator in the considered single-particle basis.
The third and fourth terms in Eq.~(\ref{eq:ham-bf}) are the exchange
interactions. They are  introduced to 
account for the fact that bosons are built from nucleon pairs and are
given by \cite{alonso1984,arias1986}
\begin{eqnarray}
\label{Rayner-new-label}
 \hat V_{\pi\nu} =&& -(s_{\pi}^\+\tilde d_{\pi})^{(2)}
\cdot
\Bigg\{
\sum_{jj'j''}
\sqrt{\frac{10}{N_\nu(2j+1)}}\beta_{jj'}\beta_{j''j} \nonumber \\
&&:((d_{\nu}^\+\times\tilde a_{j''\nu})^{(j)}\times
(a_{j\nu}^\+\times\tilde s_\nu)^{(j')})^{(2)}:
\Bigg\} + (H.c.), \nonumber \\
\end{eqnarray}
with a similar expression for  $\hat V_{\nu\pi}$.
In Eq.~(\ref{Rayner-new-label}), $\beta_{jj'}=(u_jv_{j'}+v_ju_{j'})Q_{jj'}$.
The fifth and sixth terms in Eq.~(\ref{eq:ham-bf}) are the monopole
interactions with $\hat n_{d\nu}$ and $\hat n_{d\pi}$ the number
operators for neutron and proton $d$ bosons, respectively, while the
number operator for the odd fermion $\hat
n_{\tau}=\sum_j(-\sqrt{2j+1})(a^\+_{j\tau}\times\tilde 
a_{j\tau})^{(0)}$. 

The boson-fermion Hamiltonian $\hat H_{\rm BF}$ in Eq.~(\ref{eq:ham-bf})
has been justified from microscopic considerations based on the
generalized seniority scheme \cite{IBFM,scholten1985}. 
Both the quadrupole dynamic and exchange terms act predominantly
between protons and neutrons (i.e., between odd neutron and proton
bosons and between odd proton and neutron bosons) \cite{alonso1984},
while the monopole interaction acts between like-particles 
(i.e., between odd neutron and neutron bosons and between odd proton and
proton bosons) \cite{IBFM}. 

The single-particle energies $\epsilon_{j\tau}$ and occupation 
probabilities $v^2_j$ of the odd nucleon at the  $j$ orbital are 
obtained from Gogny-D1M HFB calculations constrained to quadrupole 
moment zero. Note that this is a standard HFB calculation without 
blocking. However, the particle number is constrained to odd $N$ or 
$Z$. For more details, see Ref.~\cite{nomura2017odd-2}.

The coupling constants $\Gamma_\nu$, $\Gamma_\pi$, $\Lambda_\nu$,
$\Lambda_\pi$, $A_\nu$, and $A_{\pi}$ in Eq.~(\ref{eq:ham-bf}) are treated as free
parameters. They have been  
fitted to reproduce the energies of
the lowest-lying states in each of the odd-mass nuclei, separately for
normal-parity and unique-parity configurations. Those 
parameters have been  plotted in  Figs.~\ref{fig:para-ptos} (odd-$N$ Pt
and Os nuclei) and \ref{fig:para-ir} (odd-$Z$ Ir nuclei). 
As can be seen in
Figs.~\ref{fig:para-ptos}(a) and \ref{fig:para-ptos}(b), the
$\Gamma_\nu$ values for both the normal-parity (denoted by pfh) and
unique-parity (i13) configurations stay  rather  constant as a function of
neutron number.
The $\Lambda_\nu$ values, in Figs.~\ref{fig:para-ptos}(c) and
\ref{fig:para-ptos}(d) also look rather insensitive to variations in neutron
number except for the abrupt change in the $\Lambda_\nu$ value
for the $i_{13/2}$ configuration (see, Fig.~\ref{fig:para-ptos}(d)) in going from
$^{185}$Pt to $^{187}$Pt.  As shown later in Fig.~\ref{fig:level-pt-i13}, 
the abrupt change observed is very likely a consequence of the
evolution of the low-lying positive-parity level structure when going from
one nucleus to the other. 
Furthermore, as can be seen from Figs.~\ref{fig:para-ptos}(e) and
\ref{fig:para-ptos}(f),  the monopole strength
$A_\nu$ is chosen to be zero in many of the 
studied nuclei. Nevertheless,  a relatively large
value is needed for the transitional regions, i.e., $^{189,191}$Pt and
$^{189,191}$Os. 
The $\Gamma_\pi$, $\Lambda_\pi$ and $A_\pi$ values for the odd-$Z$ Ir isotopes are
shown in Fig.~\ref{fig:para-ir}. They are 
also nearly constant or change only moderately as functions of the neutron number $N$. Since the monopole 
interaction turns out to play a
major role for most of the considered Ir nuclei, its strength parameter is
much larger in magnitude than in the case of the odd-mass Pt and Os nuclei.

In addition, we plot in Figs.~\ref{fig:spe-ptos} and \ref{fig:spe-ir} the single-particle
energies and occupation probabilities used in the present study for the
considered odd-mass nuclei. 

The  IBFM Hamiltonian of Eq.~(\ref{eq:ham}), with  parameters 
determined via the mapping procedure, has been diagonalized to obtain 
excitation spectra and electromagnetic transition rates. For the later,  
E2 and M1 operators similar to those in Refs 
\cite{nomura2017odd-2,nomura2017odd-3} have been used. The effective 
bosonic charge  has been taken to be the same for both protons and 
neutrons, with a value $e_{\rm B}^\nu=e_{\rm B}^\pi=0.15\,e$b fitted to 
reproduce the experimental $B(E2;2^+_1\rightarrow 0^+_1)$ transition 
rate in $^{196}$Pt. For the fermion effective charges we have used the 
values $e_{\rm F}^\nu=0.3\,e$b and $e_{\rm F}^{\pi}=0.5\,e$b. Moreover, 
for the bosonic $g$-factor we have also taken the same value for 
protons and neutrons $g_{\rm B}^\nu=g_{\rm B}^\pi=0.3\,\mu_N$ that is 
fitted to reproduce the magnetic moment of the $2^+_1$ state in 
$^{196}$Pt. For the fermionic $g$-factors, we have adopted 
$g_l=1.0\,\mu_N$ for the odd proton and $g_l=0\,\mu_N$ for the odd 
neutron. The free values of $g_s$ have been quenched by 30 \%.

\section{Results for spectroscopic properties\label{sec:sys}}

In this section, we discuss the results of this study. First, in 
Sec.~\ref{sec:even}, we briefly discuss the results obtained for 
even-even nuclei. The systematics of the 
low-lying yrast levels in the odd-mass nuclei is presented in
Sec.~\ref{sec:odd}. A more detailed analysis of the level
schemes and electromagnetic properties for some selected odd-mass nuclei 
is presented in Sec.~\ref{sec:detail}.
Finally, in Sec.~\ref{sec:def}, we consider 
effective $\beta$ and $\gamma$ deformation parameters as another 
signature of the
prolate-to-oblate shape transition.

\subsection{Even-even nuclei\label{sec:even}}


\begin{figure}[htb!]
\begin{center}
\includegraphics[width=\linewidth]{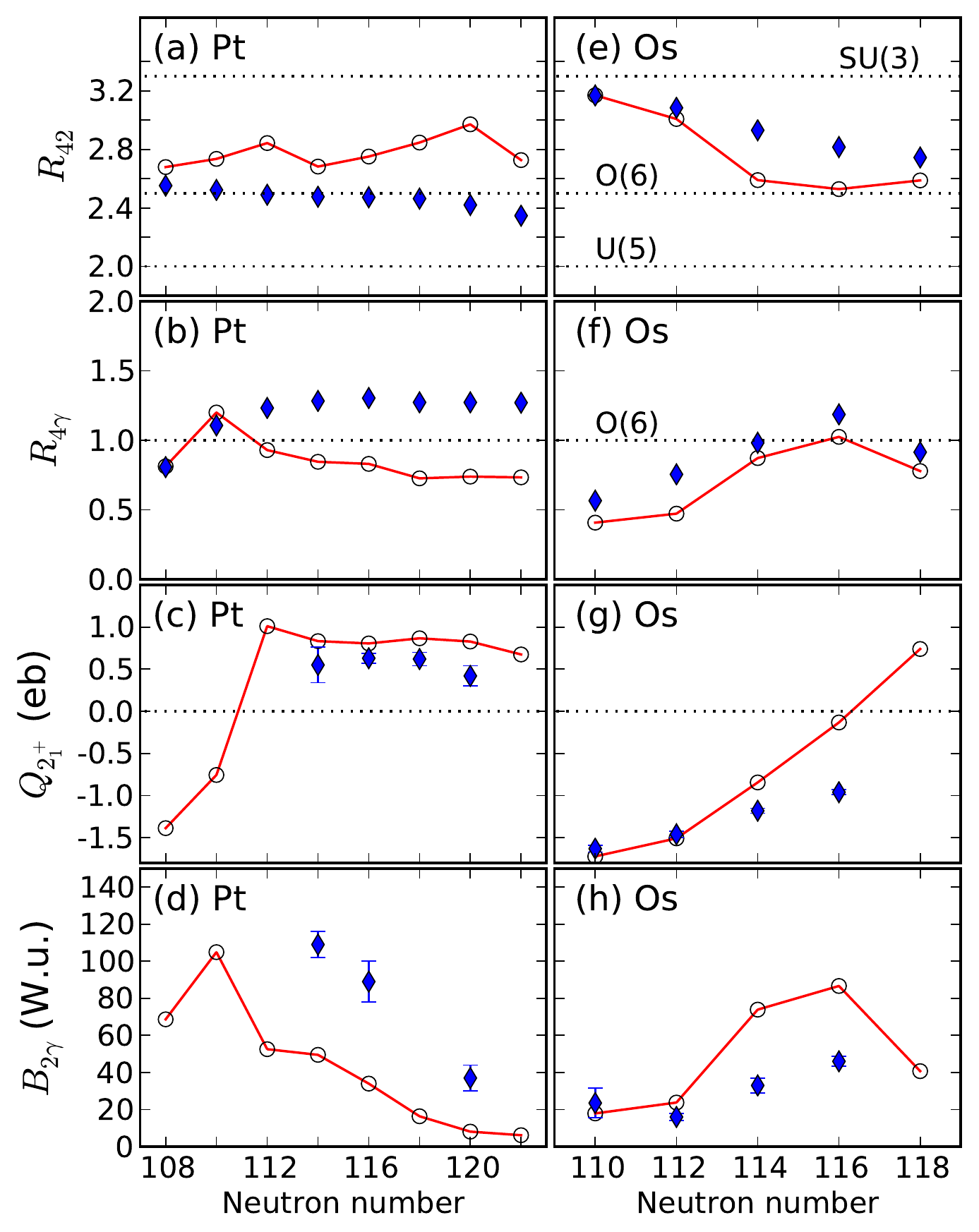}
\caption{(Color online) Spectroscopic properties of the even-even
 $^{186-200}$Pt and $^{186-194}$Os nuclei plotted as functions of the
 neutron number: the energy ratios
 $R_{42}$ and $R_{4\gamma}$, spectroscopic quadrupole
 moment  $Q_{2^+_1}$ and the $B(E2)$ value
 $B_{2\gamma}$. For more details, see the main
 text. Open circles, connected by lines, represent the theoretical values
 while solid diamonds represent the experimental data taken from
 Refs.~\cite{data,stone2005}. The symmetry limits $R_{42}=2.0$ (U(5)), 
2.5 (O(6)) and 3.3 (SU(3)), and $R_{4\gamma}=1.0$ (O(6)) of the IBM 
\cite{IBM} are also indicated.} 
\label{fig:even}
\end{center}
\end{figure}

Let us first consider how the IBM-2 Hamiltonian, extracted from the 
Gogny-D1M HFB calculations via the mapping procedure, describes the 
spectroscopic properties of the even-even nuclei. To this end, in 
Fig.~\ref{fig:even}, we have plotted spectroscopic properties of  
$^{186-200}$Pt and $^{186-194}$Os as functions of the neutron number 
$N$. The symmetry limits $R_{42}=2.0$ (U(5)), 2.5 (O(6)) and 3.3 
(SU(3)) and $R_{4\gamma}=1.0$ (O(6)) of the IBM \cite{IBM}, 
are also indicated in the figure. As can be seen in 
Fig.~\ref{fig:even}(a), 
both the theoretical and experimental $R_{42}=E(4^+_1)/E(2^+_1)$ ratios 
for Pt isotopes do not change too much and are located around the O(6) limit 
$R_{42}\approx 2.5$. On the other hand, from Fig.~\ref{fig:even}(a), 
one observes that the theoretical ratio $R_{42}$ systematically 
overestimates the experimental one for the heavier isotopes. This might be 
a consequence of the pronounced ground state deformation  in the 
corresponding Gogny-D1M energy surfaces when approaching the neutron 
closed shell $N=126$. As a result, the IBM-2 model provides a more 
rotational-like spectrum. The $R_{42}$, displayed in 
Fig.~\ref{fig:even} for Os isotopes, changes rather fast  with $N$ 
compared to the Pt isotopes. The energy ratio 
$R_{4\gamma}=E(4^+_1)/E(2^+_2)$, depicted in Figs.~\ref{fig:even}(b) 
and \ref{fig:even}(f), can be regarded as a  signature of 
$\gamma$-softness. The predicted $R_{4\gamma}$ values for both Pt and 
Os nuclei exhibit a peak at around $^{188}$Pt ($N=110$) and $^{192}$Os 
($N=116$), being close to the O(6) limit of $R_{4\gamma}=1.0$. This 
indicates that those nuclei represent the most $\gamma$-soft among the 
considered systems. Indeed, the  Gogny-D1M energy surface for 
$^{188}$Pt exhibits the most pronounced triaxial minimum at 
$\gamma\approx 30^{\circ}$ (see, Fig.~\ref{fig:pes}). As can be seen 
from Fig.~\ref{fig:even}(b), both theoretically and experimentally, the 
ratio $R_{4\gamma}$ remains rather constant in the case of Pt isotopes. 
However, the experimental $R_{4\gamma}$ values are systematically 
underestimated for $N\geq 112$. This may be due to a similar reason as 
with the discrepancy in $R_{42}$  already mentioned.

The spectroscopic quadrupole moment $Q_{2^+_1}$ for the $2^+_1$ state,
displayed in Figs.~\ref{fig:even}(c) and \ref{fig:even}(g), represents
a useful measure of whether
the nucleus is prolate or oblate. As can be seen in Fig.~\ref{fig:even}(c),
the theoretical $Q_{2^+_1}$ value is negative 
for $^{186,188}$Pt while it is positive for $^{190-200}$Pt. This is 
consistent with the prolate-to-oblate shape transition observed at the 
mean-field level in Fig.~\ref{fig:pes}. 
A similar observation can be made for the Os isotopes in Fig.~\ref{fig:even}(g).
Furthermore, the $B_{2\gamma}=B(E2; 2^+_2\rightarrow 2^+_1)$ transition
probability provides a stringent test for
$\gamma$-softness. For both Pt and Os isotopes, it shows a
peak at $N=110$ (Fig.~\ref{fig:even}(d)) and 116 (Fig.~\ref{fig:even}(h)).

\subsection{Systematics of low-energy excitation spectra in odd-mass nuclei\label{sec:odd}}


\begin{figure}[htb!]
\begin{center}
\includegraphics[width=\linewidth]{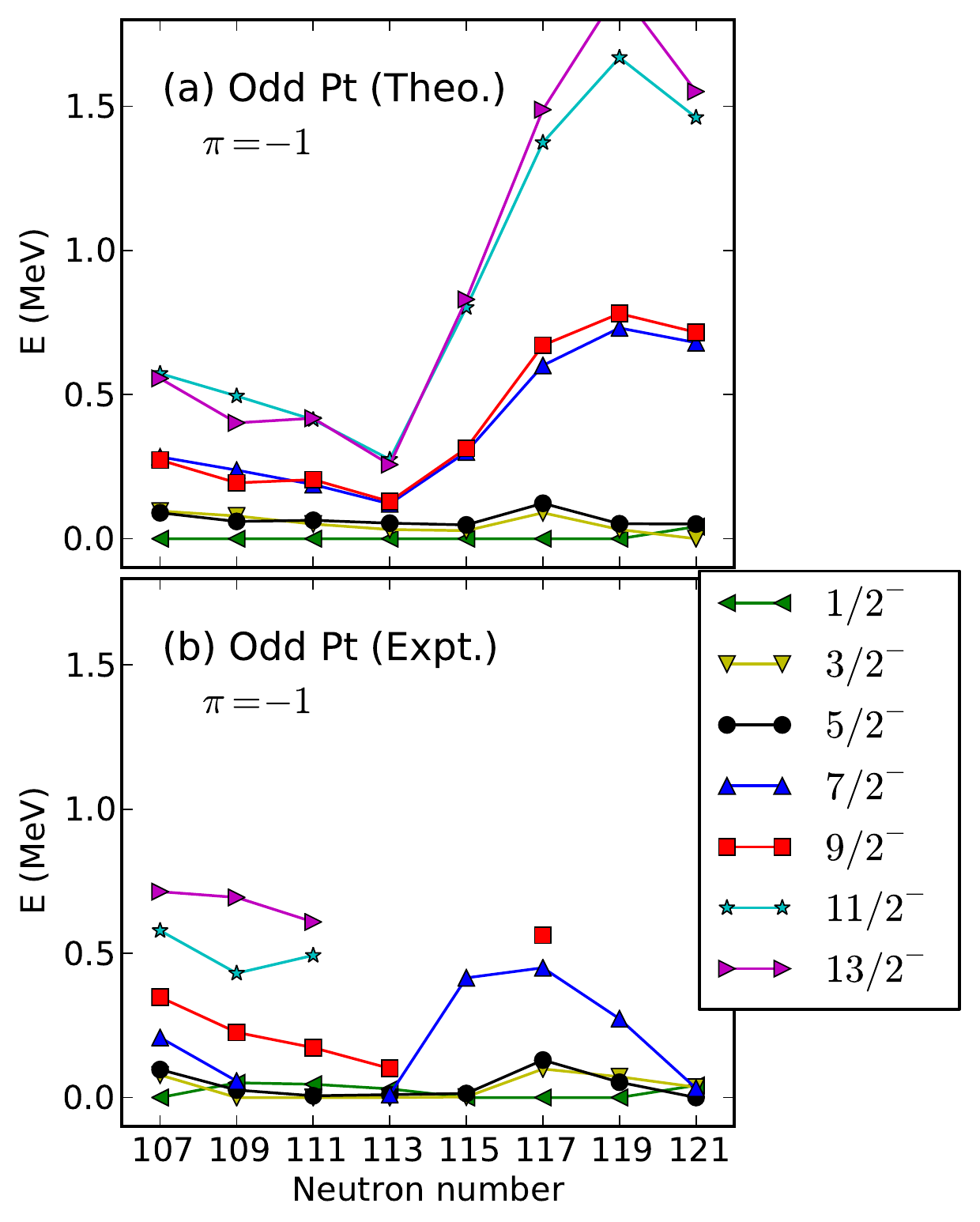}
\caption{(Color online) The theoretical and experimental low-lying
negative-parity ($\pi=-1$) yrast states in the
odd-$N$ isotopes $^{185-199}$Pt are plotted as 
functions of the neutron number.} 
\label{fig:level-pt-pfh}
\end{center}
\end{figure}


\begin{figure}[htb!]
\begin{center}
\includegraphics[width=\linewidth]{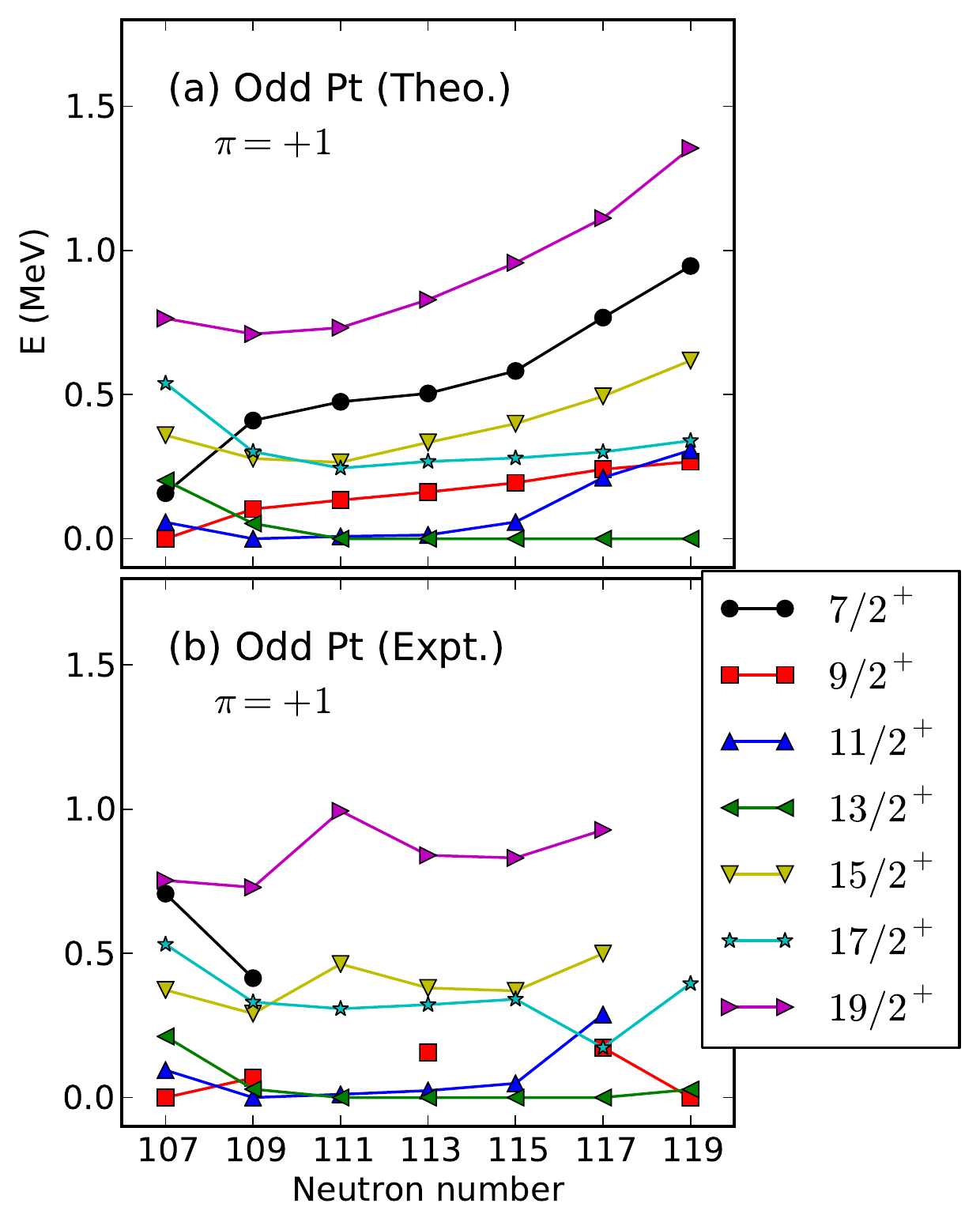}
\caption{(Color online) The same as in Fig.~\ref{fig:level-pt-pfh} but for the
positive-parity states.} 
\label{fig:level-pt-i13}
\end{center}
\end{figure}


\begin{figure}[htb!]
\begin{center}
\includegraphics[width=\linewidth]{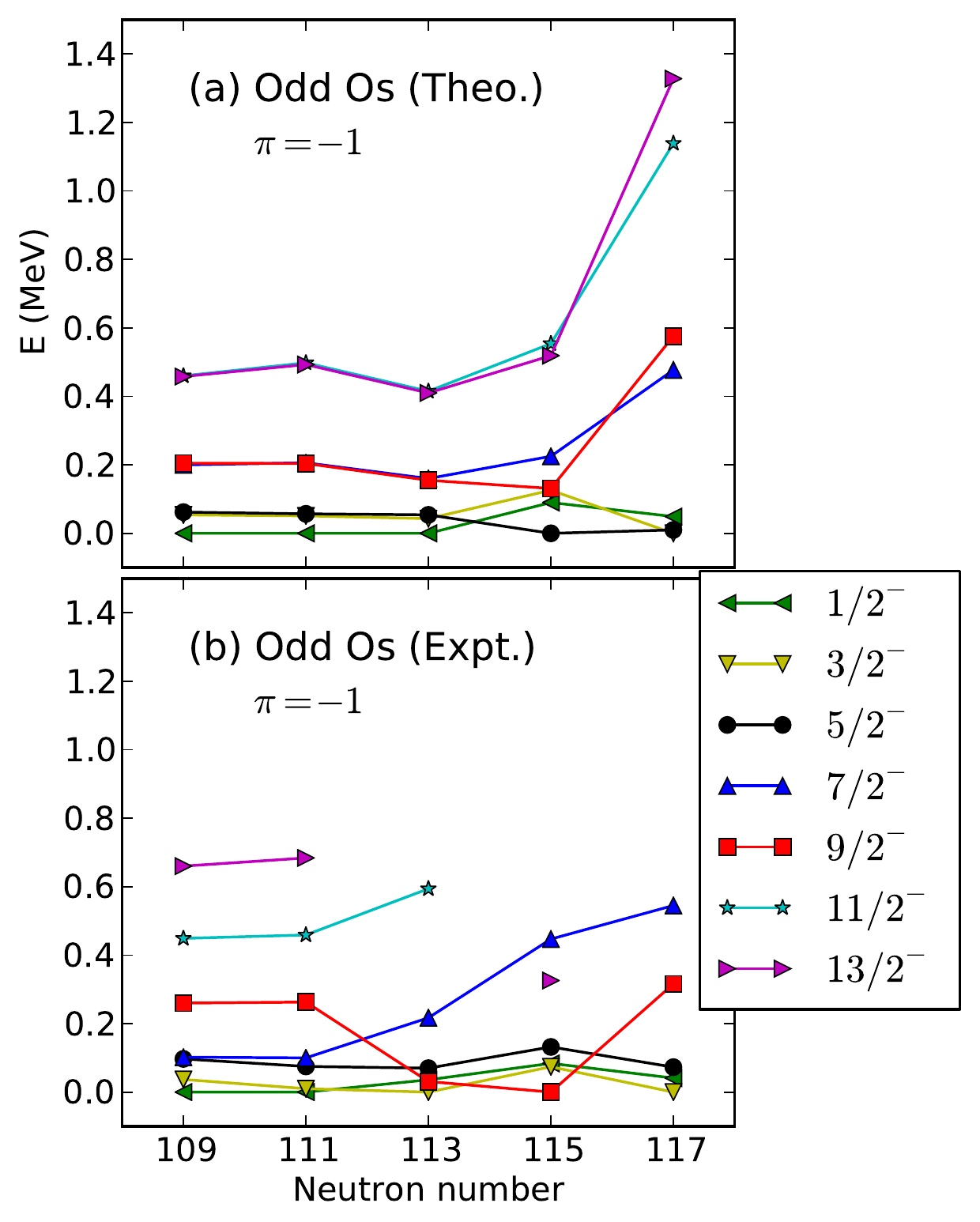}
\caption{(Color online) The  same as in Fig.~\ref{fig:level-pt-pfh}, 
but for the isotopes $^{185-193}$Os.} 
\label{fig:level-os-pfh}
\end{center}
\end{figure}


\begin{figure}[htb!]
\begin{center}
\includegraphics[width=\linewidth]{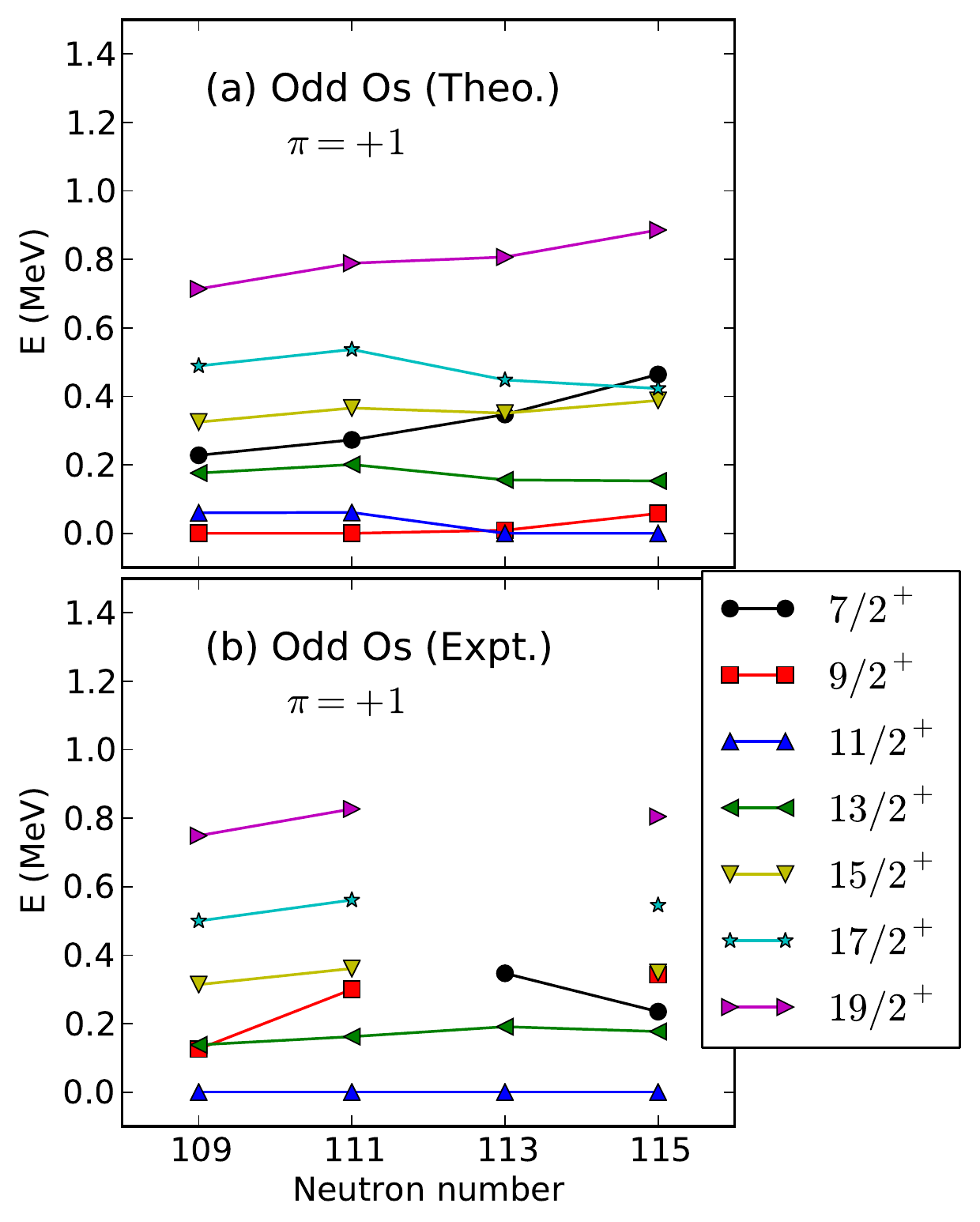}
\caption{(Color online) The same as in Fig.~\ref{fig:level-pt-i13} 
but for the isotopes $^{185-193}$Os.} 
\label{fig:level-os-i13}
\end{center}
\end{figure}


\begin{figure}[htb!]
\begin{center}
\includegraphics[width=\linewidth]{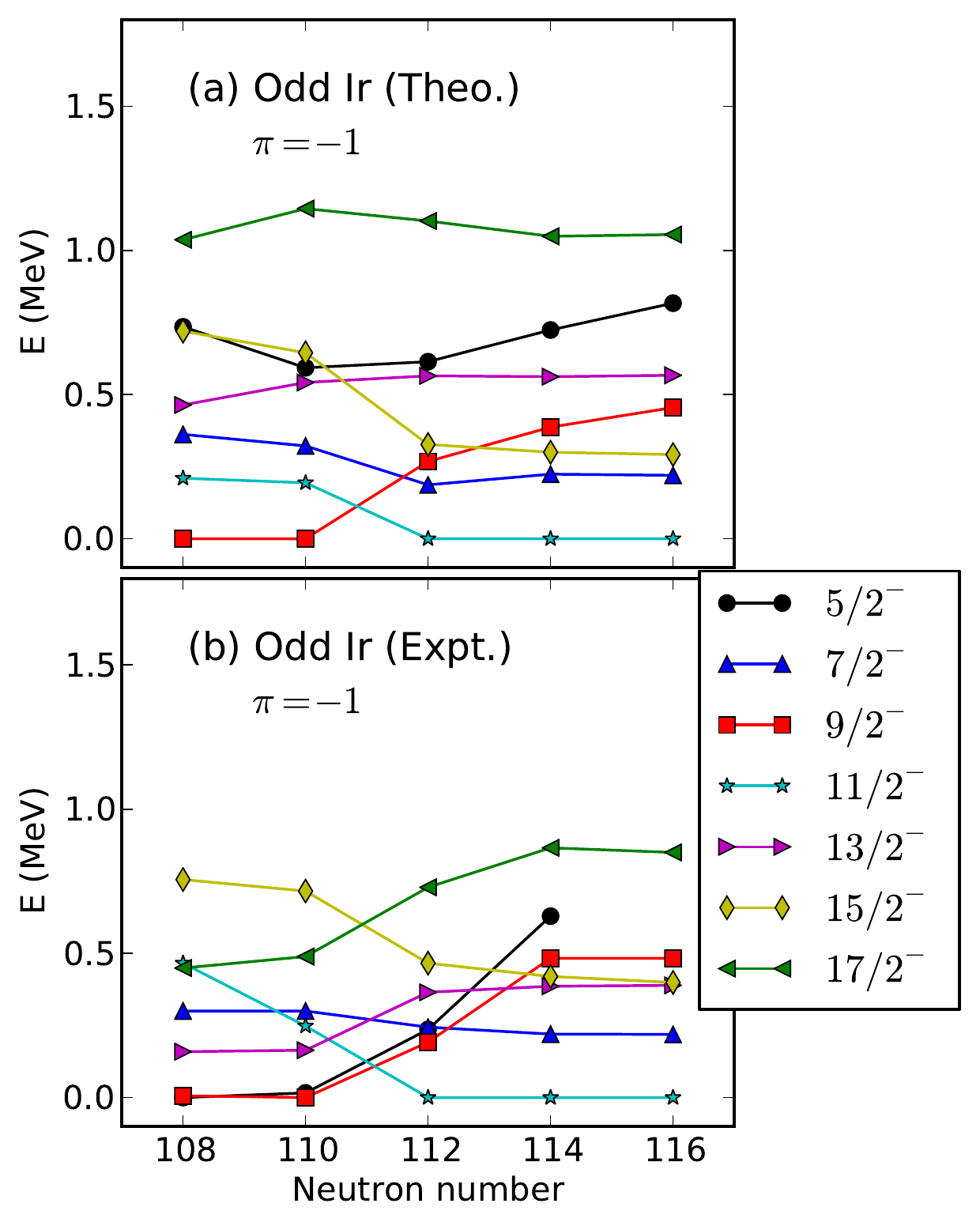}
\caption{(Color online) The same as in Fig.~\ref{fig:level-pt-pfh} 
but for the isotopes $^{185-195}$Ir.} 
\label{fig:level-ir-h11}
\end{center}
\end{figure}


\begin{figure}[htb!]
\begin{center}
\includegraphics[width=\linewidth]{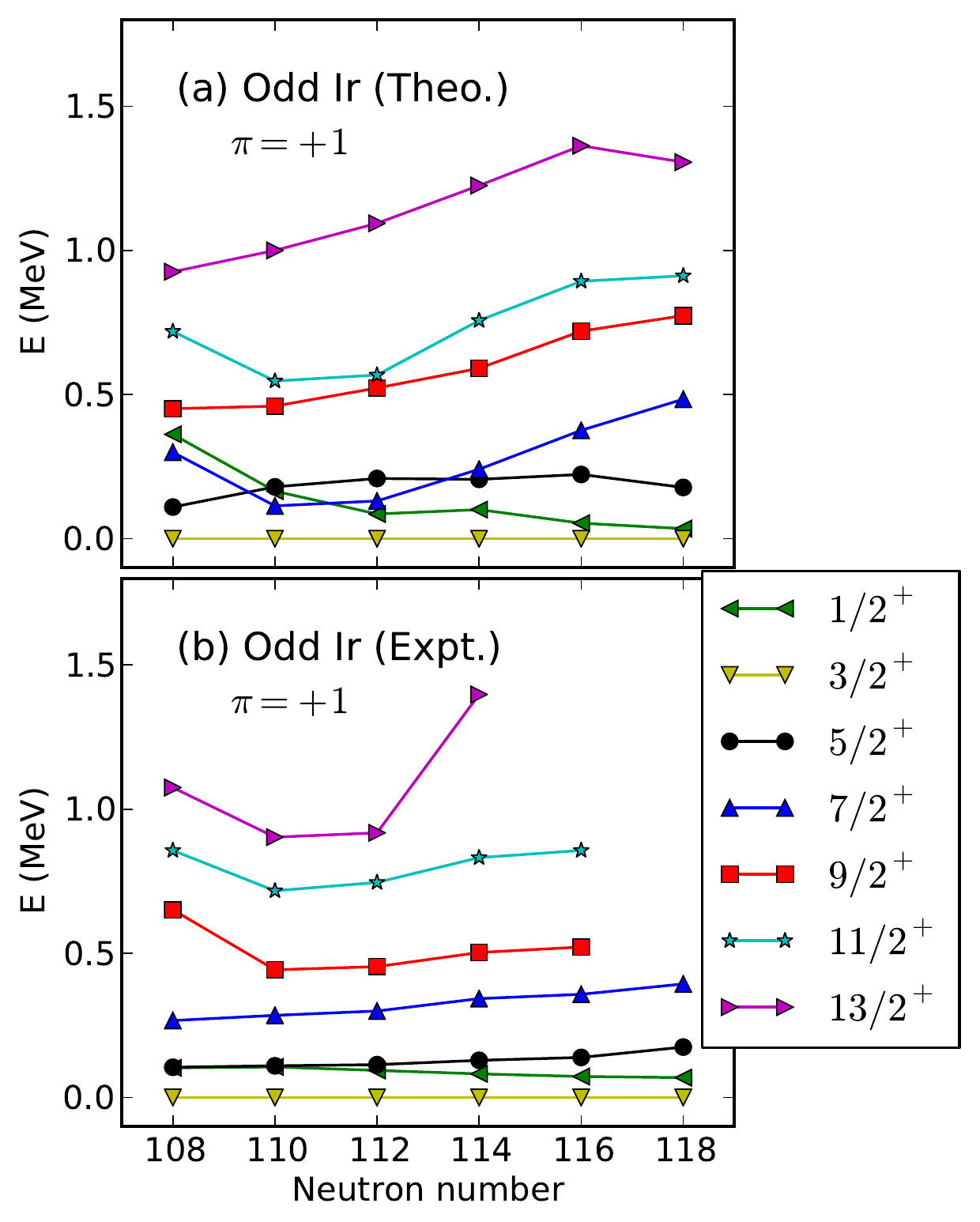}
\caption{(Color online) The same as in Fig.~\ref{fig:level-pt-i13} 
but for the isotopes $^{185-195}$Ir.} 
\label{fig:level-ir-sdg}
\end{center}
\end{figure}

In Figs.~\ref{fig:level-pt-pfh} to \ref{fig:level-ir-sdg}, we have 
plotted the energy systematics of the low-lying positive- and 
negative-parity yrast states in the odd-$N$ nuclei $^{185-199}$Pt, 
$^{185-193}$Os and $^{185-195}$Ir as functions of $N$. For the three 
isotopic chains, our calculations  describe reasonably well the 
experimental trend. For the odd-$N$ Pt isotopes, in 
Figs.~\ref{fig:level-pt-pfh} and \ref{fig:level-pt-i13}, the observed 
change in the ground state's spin in going from $N=107$ to 109 for both 
parities, can be regarded as a signature of  structural evolution and 
correlates well with the shape transition that occurs in the 
corresponding even-even systems. Indeed, the Gogny-D1M energy surfaces 
(see, Fig.~\ref{fig:pes}) suggest the transition from prolate 
($^{186}$Pt) to triaxial shapes ($^{188}$Pt). For $N=109-113$, both 
theoretically and experimentally as well as for both parities, a 
similar low-lying level structure is observed. However, as seen from 
Fig.~\ref{fig:level-pt-pfh}(a), another signature of the shape 
transition appears in the case of the negative-parity states for 
odd-$N$ Pt isotopes, i.e., at the neutron number $N=113$ many states 
are found below 0.3 MeV excitation energy while those levels higher 
than the $J={5/2}^-$ one go up rapidly for larger $N$. This also 
correlates well with the  Gogny-D1M energy surfaces obtained for 
even-even nuclei (see, Fig.~\ref{fig:pes}) which exhibit a gradual 
change of the global minimum from shallow triaxial ($^{192}$Pt) to 
oblate  ($^{194}$Pt).

The results obtained for odd-$N$ Os isotopes are shown in 
Figs.~\ref{fig:level-os-pfh} and \ref{fig:level-os-i13}. For the 
negative-parity states, in Fig.~\ref{fig:level-os-pfh}, both 
experimentally and theoretically the low-lying level structure below 
0.3 MeV excitation energy changes significantly  around $N=115$, 
including a change in the ground state's spin. Once more, this agrees 
well with the Gogny-D1M energy surfaces (see, Fig.~\ref{fig:pes}) 
suggesting a transition from a triaxial shape at $^{192}$Os ($N=116$) 
to an oblate-soft one at $^{194}$Os ($N=118$). At $N=109$ and 111, the predicted 
ground-state spins for positive parity states 
(Fig.~\ref{fig:level-os-i13}(a)) do not coincide with the experiment 
(Fig.~\ref{fig:level-os-i13}(b)). This results from the fact, that the 
boson-fermion parameters for those nuclei have been chosen so as to 
reproduce the overall level structure up to the spin $J\sim{19/2}^+$. 
However, we have also verified that if one attempts to reproduce the 
experimental ground-state's spin for $^{185,187}$Os, the whole spectrum 
becomes too compressed.

Similar observations apply to the results for the odd-$Z$ Ir nuclei, 
depicted in Figs.~\ref{fig:level-ir-h11} and \ref{fig:level-ir-sdg}. 
For example,  both the theoretical and experimental negative-parity 
spectra in Figs.~\ref{fig:level-ir-h11}(a) and 
\ref{fig:level-ir-h11}(b), respectively, suggest a rapid structural 
change in going from $N=110$ to 112. At those neutron numbers, the 
corresponding even-even Pt core nuclei undergo a structural change in 
their energy surfaces and spectroscopic properties (see, Figs.~\ref{fig:pes} and \ref{fig:even}).

\subsection{Detailed level schemes for selected odd-mass nuclei\label{sec:detail}}


\begin{figure}[htb!]
\begin{center}
\includegraphics[width=\linewidth]{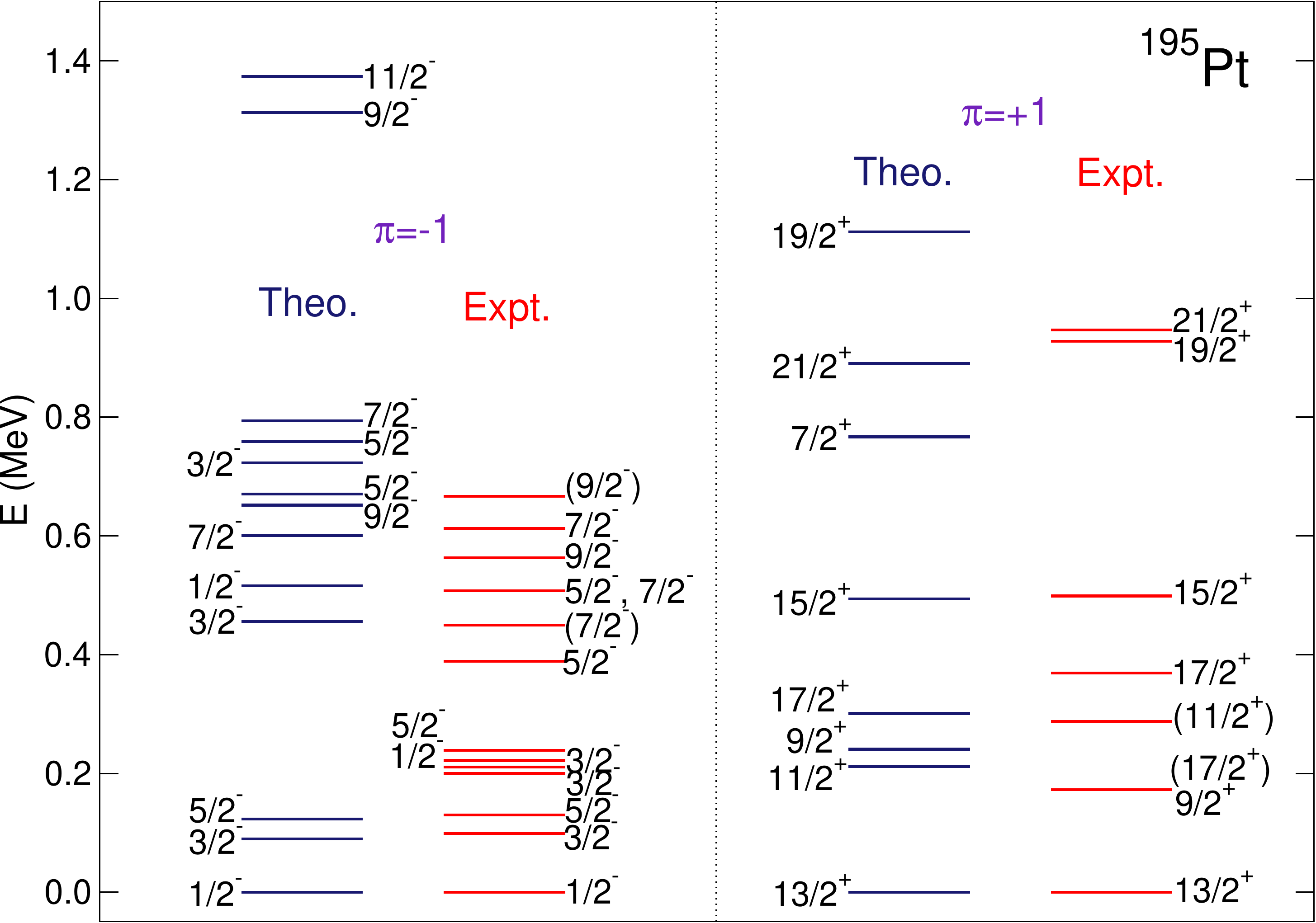}
\caption{(Color online) Comparison between the theoretical and
experimental \cite{data} low-lying positive- and negative-parity 
spectra for  $^{195}$Pt.} 
\label{fig:195pt}
\end{center}
\end{figure}


\begin{figure}[htb!]
\begin{center}
\includegraphics[width=\linewidth]{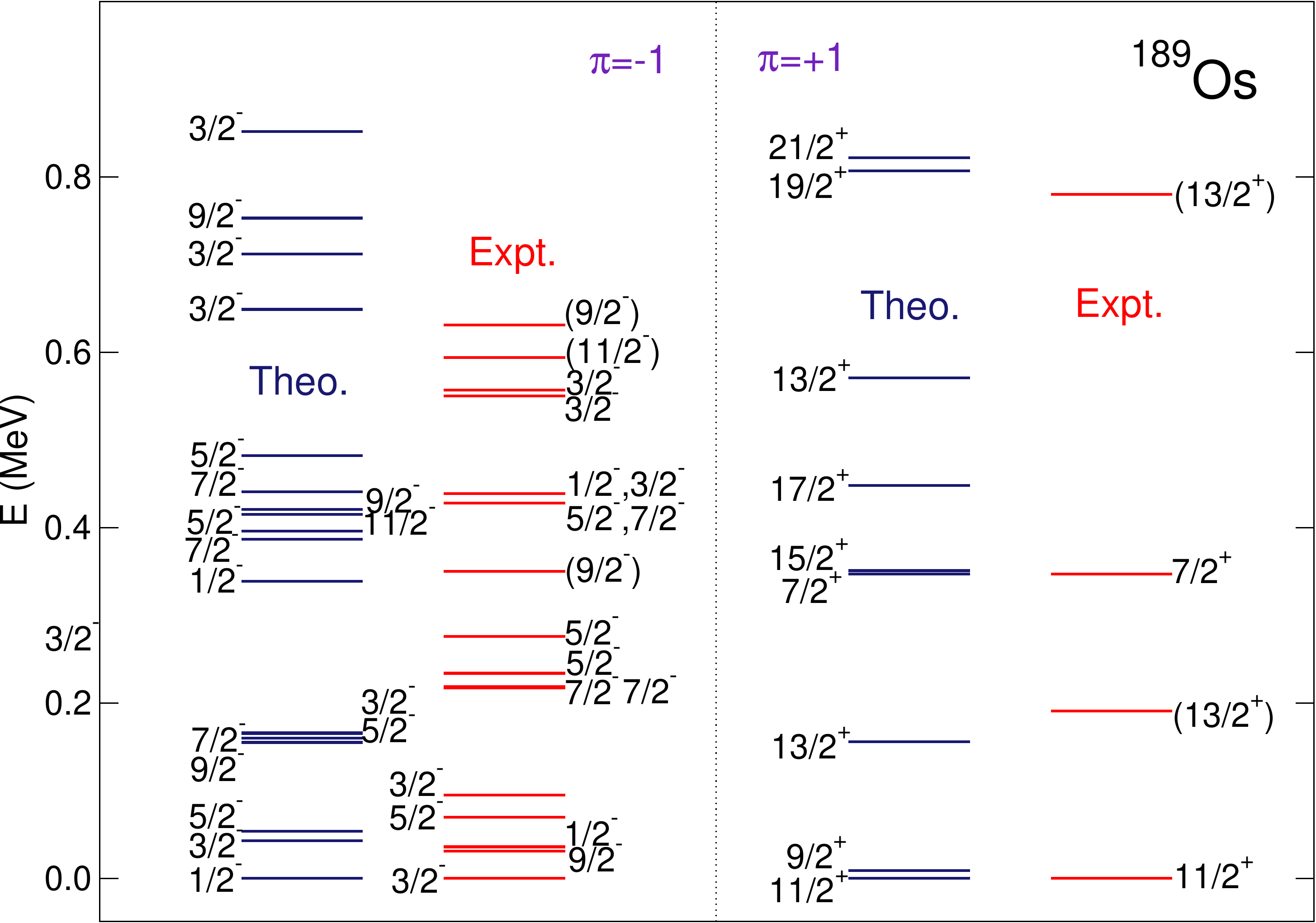}
\caption{(Color online) The same as in Fig.~\ref{fig:195pt} 
but for $^{189}$Os.} 
\label{fig:189os}
\end{center}
\end{figure}


\begin{figure}[htb!]
\begin{center}
\includegraphics[width=\linewidth]{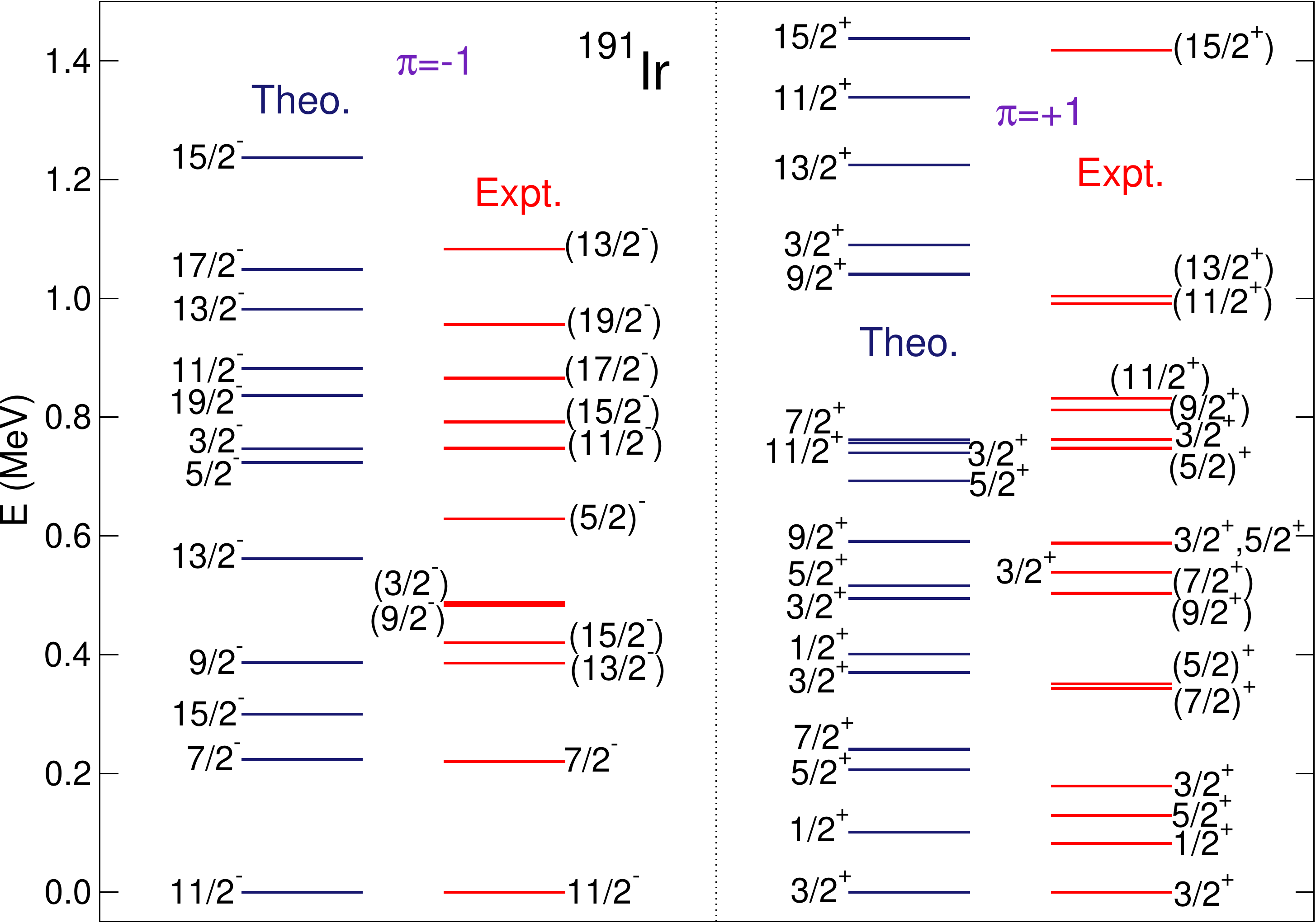}
\caption{(Color online) The same as in Fig.~\ref{fig:195pt}
but for $^{191}$Ir.} 
\label{fig:191ir}
\end{center}
\end{figure}

In this section, we present a more detailed analysis of the nuclei 
$^{195}$Pt, $^{189}$Os and $^{191}$Ir, taken as illustrative examples. 
For them, abundant experimental information, especially for 
electromagnetic properties, is available for a detailed comparison with 
the theory predictions. 

As can be seen from Fig.~\ref{fig:195pt}, our calculations reproduce 
reasonably well the experimental negative-parity yrast states for 
$^{195}$Pt. However, the predicted  non-yrast levels tend to be 
overestimated like, for example, those experimental levels around 
$\approx$0.2 MeV excitation energy. The discrepancies occur mainly 
because the single-particle energies and $v^2_j$ values used in the 
calculations may not be realistic enough to reproduce those levels. On 
the other hand, the agreement between the theoretical and experimental 
positive-parity levels is reasonable. 

In Table~\ref{tab:195pt}, we compare the predicted and experimental 
transition rates $B(E2)$ and $B(M1)$ as well as the spectroscopic 
quadrupole $Q_{J}$ and magnetic $\mu_J$ moments. The overall agreement 
is reasonably good. However, there are also some noticeable 
discrepancies. For instance, both the  $B(E2; 
{5/2}^-_2\rightarrow{3/2}^-_1)$ and $B(M1; 
{5/2}^-_2\rightarrow{3/2}^-_1)$ values are significantly smaller than 
the experimental ones. The dominant components of the  IBFM-2 wave 
function for the ${5/2}^-_2$ state are $3p_{3/2}$ (39 \%) and 
$2f_{5/2}$ (45 \%), while those for the ${3/2}^-_1$ state are 
$3p_{1/2}$ (45 \%) and $3p_{3/2}$ (37 \%). Such a  difference in the 
composition of the wave functions result in a small overlap between the
two states that is the main responsible of the too small 
E2 and M1 transition rates predicted.


\begin{table}[htb]
\caption{\label{tab:195pt}%
The theoretical $B(E2)$ and $B(M1)$ transition probabilities  (in
Weisskopf units) and the $Q_J$ (in $e$b units) and $\mu_J$ (in $\mu_N$ 
units) values for $^{195}$Pt are compared with the available
experimental data \cite{data,stone2005}. }
\begin{center}
\begin{tabular}{p{2.5cm}cccc}
\hline\hline
\multirow{2}{*}{} & \multicolumn{2}{c}{$B(E2)$ (W.u.)} &
 \multicolumn{2}{c}{$B(M1)$ (W.u.)} \\
\cline{2-3} 
\cline{4-5}
          & Th.         & Exp.    & Th.         & Exp.     \\
\hline
${3/2}^-_1\rightarrow {1/2}^-_1$ & 24 & 11.5(15) & 0.034 & 0.0168(19) \\
${3/2}^-_2\rightarrow {1/2}^-_1$ & 1.8 & 4.5(13) & 0.0066 & 0.00033(11) \\
${3/2}^-_3\rightarrow {1/2}^-_1$ & 7.8 & 30(7) & 0.029 & 0.024(4) \\
${3/2}^-_4\rightarrow {1/2}^-_1$ & 2.2 & 0.22(7) & 0.0060 & 0.0036(7) \\
${5/2}^-_1\rightarrow {1/2}^-_1$ & 21 & 8.9(7) & - & - \\
${5/2}^-_2\rightarrow {1/2}^-_1$ & 14 & 49(7) & - & - \\
${5/2}^-_3\rightarrow {1/2}^-_1$ & 0.00043 & 1.3(9) & - & - \\
${3/2}^-_4\rightarrow {1/2}^-_2$ & 2.2 & $<$37 & 0.0083 & $>$0.00054 \\
${3/2}^-_2\rightarrow {3/2}^-_1$ & 15 & 0.05$^{+106}_{-5}$ & 0.014 & 0.0030(8) \\
${3/2}^-_4\rightarrow {3/2}^-_1$ & 4.1 & 0.07(6) & 0.027  & 0.0013(3) \\
${5/2}^-_1\rightarrow {3/2}^-_1$ & 10 & 4.8(19) & 0.012 & 0.0269(21) \\
${5/2}^-_2\rightarrow {3/2}^-_1$ & 0.082 & 11(6) & 0.0015 & 0.019(3) \\
${5/2}^-_3\rightarrow {3/2}^-_1$ & 11 & 38(20) & 0.041 & 0.038(17) \\
${5/2}^-_4\rightarrow {3/2}^-_1$ & - & - & 0.00050 & $<0.013$ \\
${7/2}^-_2\rightarrow {3/2}^-_1$ & 4.1 & 29(10) & - & - \\
${5/2}^-_4\rightarrow {3/2}^-_3$ & - & - & 0.0017 & $<0.017$\\
${7/2}^-_2\rightarrow {3/2}^-_3$ & 2.2 & 7(3) & - & - \\
${7/2}^-_3\rightarrow {3/2}^-_3$ & 11 & 26(17) & - & - \\
${5/2}^-_3\rightarrow {5/2}^-_1$ & 0.57 & 0.0097 & 0.0044 &
 0.026(12)\\
${7/2}^-_2\rightarrow {5/2}^-_1$ & - & - & 0.031 & 0.014(5) \\
${9/2}^-_1\rightarrow {5/2}^-_1$ & 45 & 35(8) & - & - \\
${5/2}^-_3\rightarrow {5/2}^-_2$ & - & - & 0.057 & 0.030(15) \\
${5/2}^-_4\rightarrow {5/2}^-_2$ & 0.33 & $<60$ & - & - \\
${7/2}^-_3\rightarrow {5/2}^-_2$ & 0.12 & $<210$ & 0.00077 & $<0.077$ \\
${9/2}^-_2\rightarrow {5/2}^-_2$ & 34 & 30(8) & - & - \\
${7/2}^-_2\rightarrow {5/2}^-_3$ & 0.11 & $<3.9\times 10^{3}$ & 0.00042 & $<0.14$ \\
\hline
\multirow{2}{*}{} & \multicolumn{2}{c}{$Q_J$ ($e$b)} &
 \multicolumn{2}{c}{$\mu_J$ ($\mu_N$)} \\
\cline{2-3} 
\cline{4-5}
          & Theo.         & Exp.    & Theo.         & Exp.     \\
\hline
${1/2}^-_1$ & - & - & +0.46  & +0.60952(6) \\
${3/2}^-_1$ & +0.46 & & -0.37 & -0.62(6) \\
${3/2}^-_3$ & +0.083 & & -0.59 & +0.16(3) \\
${5/2}^-_1$ & +0.75 & & +0.90 & +0.90(6) \\
${5/2}^-_2$ & +0.20 & & +1.06 & +0.52(5) \\
${5/2}^-_3$ & +0.41 & & -0.062 & +0.39(10) \\
${5/2}^-_4$ & -0.48 & & +0.87 & +1.6(6) \\
${7/2}^-_2$ & +0.50 & & +0.84 & +0.55(8) \\
${7/2}^-_3$ & +0.26 & & +0.78 & +1.4(4) \\
${7/2}^-_5$ & +0.11 & & +0.76 & +1.2(3) \\
${9/2}^-_2$ & +0.70 & & +1.53 & +1.55(12) \\
${9/2}^-_3$ & +0.76 & & +0.59 & +1.52(16) \\
${13/2}^+_1$ & +0.79 & +1.4(6) & -1.31 & -0.606(105) \\
\hline\hline
\end{tabular}
\end{center}
\end{table}

For  $^{189}$Os, as seen in Fig.~\ref{fig:189os}, the obtained  level 
structure is similar to the one  for $^{195}$Pt. However, the 
negative-parity spectrum differs from the experimental one, for 
example, with respect to the ground-state spin. In addition, the very 
low-lying ${9/2}^-_1$ level near the ground state could not be 
reproduced. Empirically, the ${9/2}^-_1$ state arises from the 
${1h_{9/2}}$ orbital coming closer to the Fermi surface \cite{data}. However, in 
the calculations  the single-particle energy for the $1h_{9/2}$ orbital 
lies much higher than all the other orbitals (see, 
Fig.~\ref{fig:spe-ptos}(c)). Let us remark, that such a feature cannot 
be controlled via three boson-fermion interaction strengths alone.

In Table~\ref{tab:189os} we compare the electromagnetic properties 
obtained for $^{189}$Os with the available experimental data 
\cite{data,stone2005}. Most of the discrepancy  is found for those 
transitions that involve non-yrast states. Note that, indeed, the 
energy levels corresponding to those states are neither properly 
reproduced.


\begin{table}[htb]
\caption{\label{tab:189os}%
The same as in Table~\ref{tab:195pt} but for $^{189}$Os. }
\begin{center}
\begin{tabular}{p{2.5cm}cccc}
\hline\hline
\multirow{2}{*}{} & \multicolumn{2}{c}{$B(E2)$ (W.u.)} &
 \multicolumn{2}{c}{$B(M1)$ (W.u.)} \\
\cline{2-3} 
\cline{4-5}
          & Theo.         & Exp.    & Theo.         & Exp.     \\
\hline
${3/2}^-_2\rightarrow {1/2}^-_1$ & 1.9 & 27(16) & 0.00044 & 0.048(8) \\
${5/2}^-_1\rightarrow {1/2}^-_1$ & 61 & 24(3) & -  & - \\
${5/2}^-_2\rightarrow {1/2}^-_1$ & 4.8 & 25$^{+5}_{-8}$ & - & - \\
${5/2}^-_3\rightarrow {1/2}^-_1$ & 0.0029 & 0.6(4) & - & - \\
${1/2}^-_1\rightarrow {3/2}^-_1$ & 134  & 27(7) & 0.0070 & 0.042(3) \\
${3/2}^-_2\rightarrow {3/2}^-_1$ & 45  & 14(3) & 0.014 & 0.0032(6) \\
${5/2}^-_1\rightarrow {3/2}^-_1$ & 7.7 & 100(10) & 0.019 & 0.0026(2) \\
${5/2}^-_2\rightarrow {3/2}^-_1$ & 10 & 10$^{+3}_{-4}$ & 0.00058 &  0.0005$^{+3}_{-4}$ \\
${5/2}^-_3\rightarrow {3/2}^-_1$ & 3.9  & 1.5(3) & 0.0021 & $8.9\times 10^{-5}$(23) \\
${7/2}^-_1\rightarrow {3/2}^-_1$ & 82 & 18.2(11) & - & - \\
${7/2}^-_2\rightarrow {3/2}^-_1$ & 0.94 & 38(2) & - & - \\
${5/2}^-_2\rightarrow {3/2}^-_2$ & 9.0  & 17$^{+6}_{-7}$ & 0.018 & 0.0012$^{+4}_{-5}$ \\
${5/2}^-_3\rightarrow {3/2}^-_2$ & 50 & 0.53(50) & 0.0013 & $<7.0\times
 10^{-5}$ \\
${7/2}^-_1\rightarrow {3/2}^-_2$ & 0.00077  & 1.75(22) & - & - \\
${7/2}^-_2\rightarrow {3/2}^-_2$ & 7.7 & 5(3) & - & - \\
${3/2}^-_2\rightarrow {5/2}^-_1$ & 13 & 80$^{+90}_{-40}$ & 0.00057  &
 0.011(7) \\
${5/2}^-_2\rightarrow {5/2}^-_1$ & 32 & $<16$ & 0.0018 & 0.00087(63) \\
${5/2}^-_3\rightarrow {5/2}^-_1$ & 0.49 & 1.05(33) & 0.0097 & $<3.3\times
 10^{-5}$ \\
${7/2}^-_1\rightarrow {5/2}^-_1$ & 10 & 14(6) & 0.00059 & 0.0008(4) \\
${7/2}^-_2\rightarrow {5/2}^-_1$ & 7.0 & 43(2) & - & - \\
${5/2}^-_3\rightarrow {7/2}^-_2$ & - & - & 0.029  & 0.0099(21) \\
${5/2}^-_3\rightarrow {9/2}^-_1$ & 1.9 & 41(8) & - & - \\
${7/2}^-_1\rightarrow {9/2}^-_1$ & 1.5 & $<$2.2 & 0.026 & 0.00107(17) \\
${7/2}^-_2\rightarrow {9/2}^-_1$ & 18 & 6$^{+2}_{-1}$ & 0.000 &
 0.00025$^{+11}_{-14}$ \\
\hline
\multirow{2}{*}{} & \multicolumn{2}{c}{$Q_J$ ($e$b)} &
 \multicolumn{2}{c}{$\mu_J$ ($\mu_N$)} \\
\cline{2-3} 
\cline{4-5}
          & Theo.         & Exp.    & Theo.         & Exp.     \\
\hline
${1/2}^-_1$ & - & - & +0.45 & +0.23(3) \\
${3/2}^-_1$ & -0.53 & +0.98(6) & +0.17 & +0.6599 \\
${5/2}^-_1$ & -1.03 & -0.63(2) & +0.96 & +0.988(6) \\
\hline\hline
\end{tabular}
\end{center}
\end{table}

In Fig.~\ref{fig:191ir} we compare the excitation spectra for 
$^{191}$Ir.  Both the positive- 
and negative-parity states are rather well described. The
electromagnetic properties computed for 
this nucleus are given in Table~\ref{tab:191ir}. Although the 
corresponding energy levels are reasonably well described, some 
transition strengths, like the $B(E2; {1/2}^+_1\rightarrow {3/2}^+_1)$ 
one, are significantly underestimated. 
The reason is 
that the IBFM-2 wave functions of the ${3/2}^+_1$ and ${1/2}^+_1$ 
states are mainly built from the $2d_{3/2}$ (54 \%) and 
$3s_{1/2}$ (58 \%) single-particle configurations, respectively. As a 
consequence, the E2 matrix element between the two states becomes too 
small. Keeping in mind that the employed model contains only three free 
parameters for each nucleus, the predicted electromagnetic properties 
in Table~\ref{tab:191ir} for $^{191}$Ir, together with those for 
$^{195}$Pt (Table~\ref{tab:195pt}) and $^{189}$Os 
(Table~\ref{tab:189os}), appear to be rather reasonable.


\begin{table}[htb]
\caption{\label{tab:191ir}%
The same as in Table~\ref{tab:195pt}, but for $^{191}$Ir.}
\begin{center}
\begin{tabular}{p{2.2cm}cccc}
\hline\hline
\multirow{2}{*}{} & \multicolumn{2}{c}{$B(E2)$ (W.u.)} &
 \multicolumn{2}{c}{$B(M1)$ (W.u.)} \\
\cline{2-3} 
\cline{4-5}
          & Theo.         & Exp.    & Theo.         & Exp.     \\
\hline
${1/2}^+_2\rightarrow {1/2}^+_1$ & - & - & 0.00096 & $<0.0038$ \\
${3/2}^+_2\rightarrow {1/2}^+_1$ & 2.4 & 58(9) & 0.00064 & 0.066(8) \\
${3/2}^+_3\rightarrow {1/2}^+_1$ & 15 & 0.39(10) & 0.00016 & 0.00204(21) \\
${5/2}^+_1\rightarrow {1/2}^+_1$ & 0.37 & 10.4(13) & - & - \\
${5/2}^+_2\rightarrow {1/2}^+_1$ & 5.5 & 39(7) & - & - \\
${1/2}^+_1\rightarrow {3/2}^+_1$ & 0.32 & 20.9(7) & 3.3$\times 10^{-6}$ & 0.000474(14) \\
${1/2}^+_2\rightarrow {3/2}^+_1$ & 13 & $<2.7$ & 0.0070 & $<0.000277$(23) \\
${3/2}^+_2\rightarrow {3/2}^+_1$ & 24 & 15.7(20) & 0.029 & 0.00205(24) \\
${3/2}^+_4\rightarrow {3/2}^+_1$ & 0.27 & 2.52(25) & 4.9$\times 10^{-5}$  & 0.0040(4) \\
${3/2}^+_5\rightarrow {3/2}^+_1$ & 0.23 & - & 8.2$\times 10^{-5}$ & $\approx 0.17$ \\
${5/2}^+_1\rightarrow {3/2}^+_1$ & 49 & 96.2(24) & 0.11 & 0.0259(6) \\
${5/2}^+_2\rightarrow {3/2}^+_1$ & 9.6 & 1.7(4) & 0.044 & 0.0060(9) \\
${5/2}^+_3\rightarrow {3/2}^+_1$ & 0.011 & - & 0.0087 & $\approx 0.17$ \\
${7/2}^+_1\rightarrow {3/2}^+_1$ & 21 & 41.9(20) & - & - \\
${7/2}^+_2\rightarrow {3/2}^+_1$ & 8.1 & 8.9(11) & - & - \\
${3/2}^+_3\rightarrow {3/2}^+_2$ & 7.2 & 0.013(10) & 0.067 & 0.0080(7) \\
${5/2}^+_2\rightarrow {3/2}^+_2$ & 1.9 & 3.9(9) & 0.0015 & 0.056(8) \\
${1/2}^+_2\rightarrow {5/2}^+_1$ & 3.4 & $<1.3$ & - & - \\
${3/2}^+_2\rightarrow {5/2}^+_1$ & 24 & 27(9) & 0.00055  & 0.0053(11) \\
${3/2}^+_3\rightarrow {5/2}^+_1$ & 4.7 & 0.60(9) & 0.00034  & 0.0071(8) \\
${7/2}^+_1\rightarrow {5/2}^+_1$ & 6.8 & 29.8(15) & 0.0071 & 0.0296(15) \\
${7/2}^+_2\rightarrow {5/2}^+_1$ & 13 & 18(3) & 0.022 & 0.0117(18) \\
${9/2}^+_1\rightarrow {5/2}^+_1$ & 33 & 72(11) & - & - \\
${3/2}^+_3\rightarrow {5/2}^+_2$ & 0.022 & 10(4) & 0.0052 & 0.0031(5) \\
${11/2}^+_1\rightarrow {7/2}^+_1$ & 23 & 70(4) & - & - \\
${3/2}^-_1\rightarrow {7/2}^-_1$ & 80 & $>38$ & - & - \\
${7/2}^-_1\rightarrow {11/2}^-_1$ & 78 & 56(5) & - & - \\
\hline
\multirow{2}{*}{} & \multicolumn{2}{c}{$Q_J$ ($e$b)} &
 \multicolumn{2}{c}{$\mu_J$ ($\mu_N$)} \\
\cline{2-3} 
\cline{4-5}
          & Theo.         & Exp.    & Theo.         & Exp.     \\
\hline
${1/2}^+_1$ & - & - & +1.20 & +0.600(6) \\
${3/2}^+_1$  & +0.32 & +0.816(9) & +0.29 & +0.1507(6) \\
${5/2}^+_1$ &  & - & +1.37 & +0.81(6) \\
${7/2}^+_1$ &  & - & +0.99 & +1.40(6) \\
${9/2}^+_1$ &  & - & +2.11 & +2.4(2) \\
${11/2}^-_1$ &  & - & +6.66 & +6.03(4) \\
\hline\hline
\end{tabular}
\end{center}
\end{table}

\subsection{Signatures of shape phase transitions\label{sec:def}}


\begin{figure}[htb!]
\begin{center}
\includegraphics[width=0.7\linewidth]{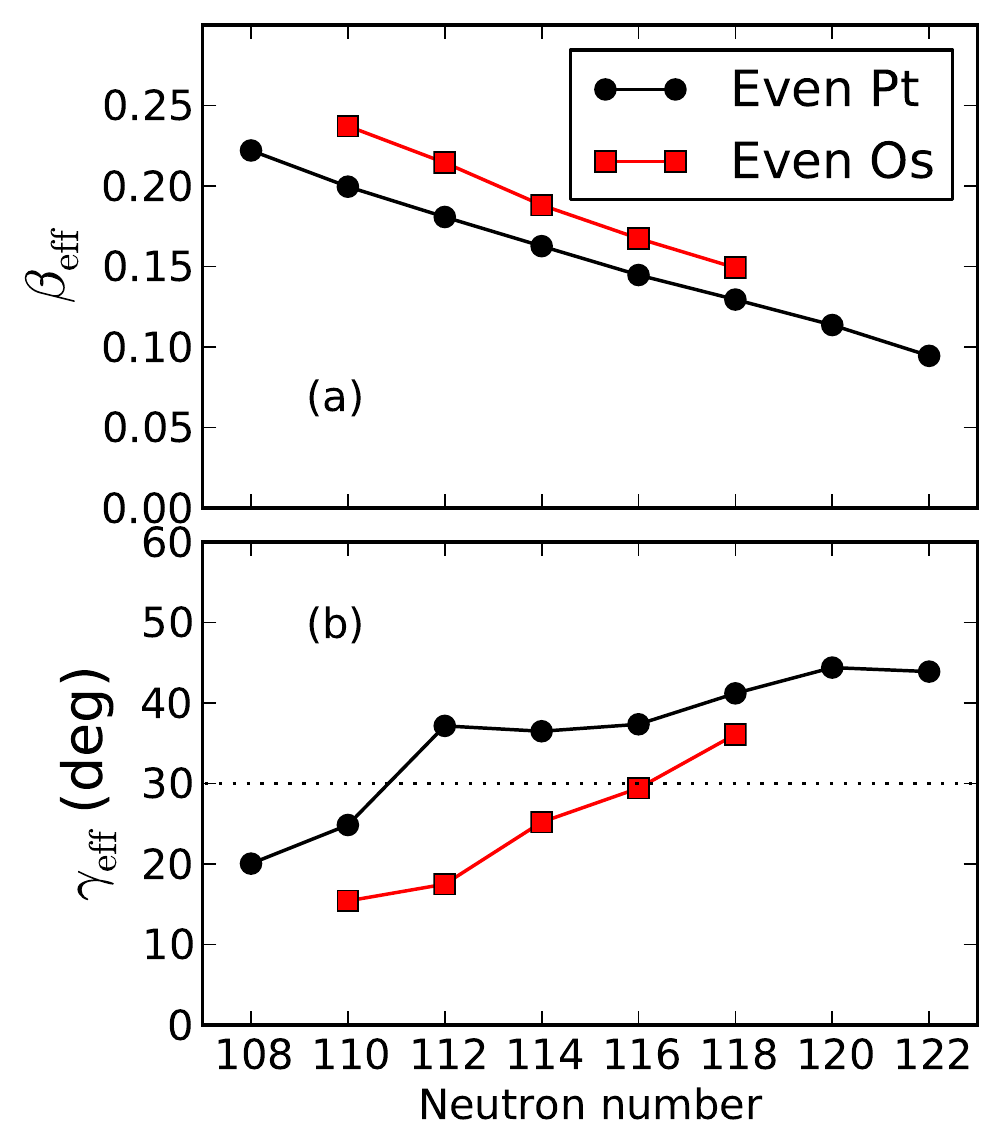}
\caption{(Color online) Effective $\beta$ and $\gamma$ deformation
parameters for the even-even nuclei $^{186-200}$Pt and $^{186-194}$Os 
obtained from the E2 transition matrix elements.} 
\label{fig:def-even}
\end{center}
\end{figure}


\begin{figure}[htb!]
\begin{center}
\includegraphics[width=\linewidth]{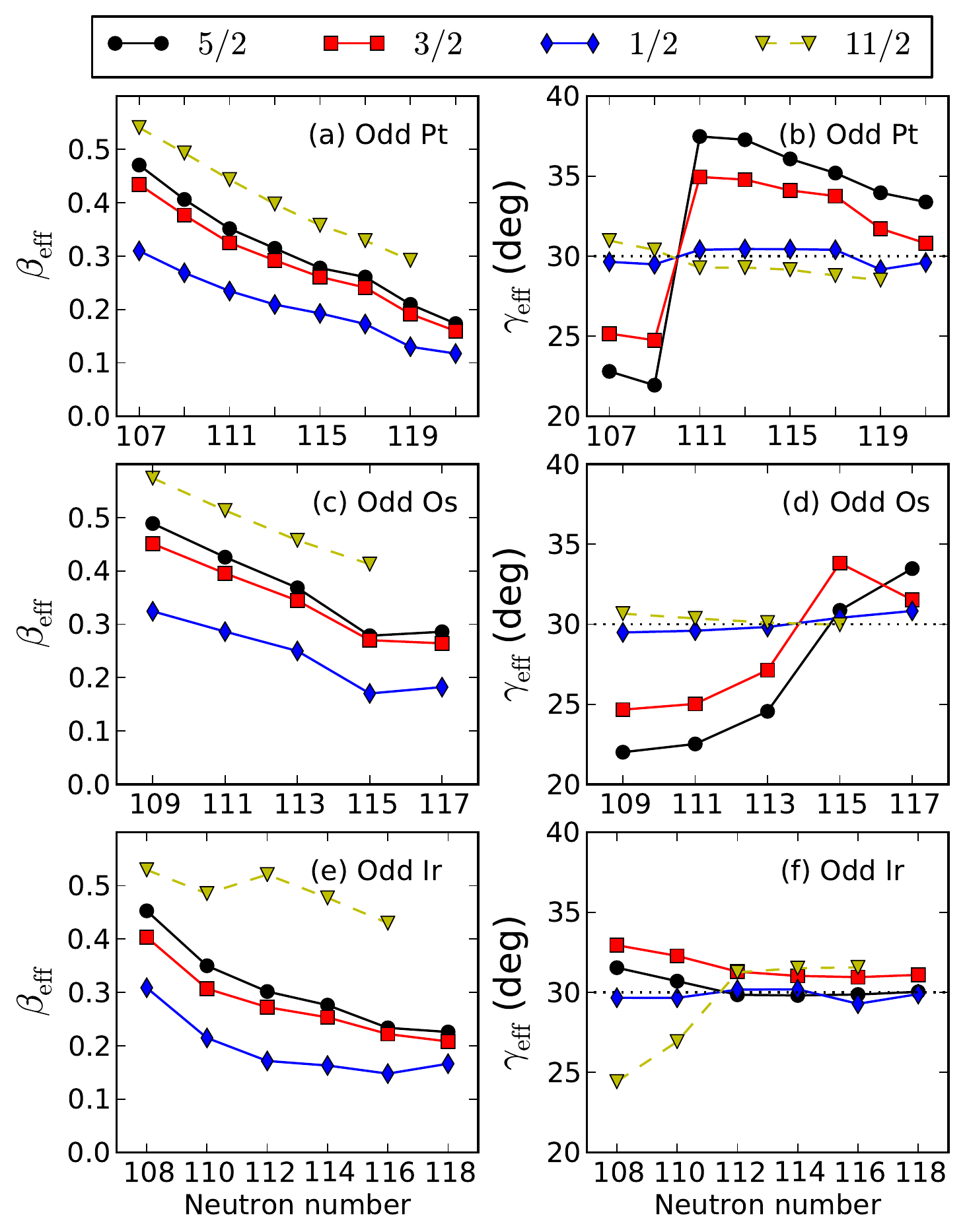}
\caption{(Color online) Effective $\beta$ and $\gamma$ deformation
 parameters for the odd-$N$ nuclei $^{185-199}$Pt and $^{185-193}$Os and
 the odd-$Z$ nuclei 
 $^{185-191}$Ir for the $J^{\pi}={1/2}^\pi_1$, ${3/2}^\pi_1$ and
 ${5/2}^\pi_1$ states for normal-parity configurations ($\pi=-1$ for Pt
 and Os, and $\pi=+1$ for Ir) and
 $J^\pi={11/2}^\pi_1$ state for unique-parity configurations ($\pi=+1$ for Pt
 and Os, and $\pi=-1$ for Ir). } 
\label{fig:def-odd}
\end{center}
\end{figure}

As yet another signature of the  prolate-to-oblate shape phase 
transitions, we consider the quadrupole shape invariants 
\cite{cline1986} (denoted as q-invariants) $q_m$ ($m=2,3,\ldots$) 
obtained from the E2 matrix elements. These quantities have already 
been shown \cite{nomura2017odd-1} to be  good signatures of shape phase 
transitions involving $\gamma$-softness.  For our purpose in this work, 
the relevant $q_m$'s read
 \begin{eqnarray}
\label{eq:q2}
q_2=\sum_{J'}q_2(J')
\end{eqnarray}
with 
\begin{eqnarray}
\label{eq:q2a}
 q_2(J')=\sum_{i}^n\langle J||\hat Q||J^{\prime}_i\rangle\langle
  J^{\prime}_i||\hat Q||J\rangle,
\end{eqnarray}
and
\begin{eqnarray}
\label{eq:q3}
q_3=-\sqrt{\frac{7}{10}}
\sum_{J'J''}
\sum_{ij}^{n}
\langle J||\hat Q||J'_i\rangle
\langle J'_i||\hat Q||J''_j\rangle
\langle J''_j||\hat Q||J\rangle.
\end{eqnarray}
In Eqs.~(\ref{eq:q2a}) and (\ref{eq:q3}), all possible E2 transition matrix elements
among the states  $J$, $J'$ and $J''$, that satisfy the E2 selection
rule, have been considered. 
The indices $i$ ($j$) in the sums are ordered according to increasing excitation
energies of the $J'$ ($J''$) levels and run up to $n=\infty$. However, we have
confirmed that only a few of the lowest transitions contribute 
to the q-invariants significantly 
\cite{nomura2017odd-1}. In the case of even-even systems, the q-invariants for the
$0^+_1$ ground state (i.e, $J=0^+_1$ and $J^{\prime}=2^+$)  have been 
computed. The effective deformation parameters
$\beta_{\rm eff}$ and $\gamma_{\rm  eff}$ are obtained
from the $q_2$ and $q_3$ values by the formulas
\begin{eqnarray}
\label{eq:eff_beta}
 &&\beta_{\rm eff} = \frac{4\pi}{3ZR^2_0}\sqrt{\sum_{J'}\frac{1}{2J^{\prime}+1}(J^{\prime}2J0|JJ)^{-2}q_2(J')} \\
\label{eq:eff_gamma}
 &&\gamma_{\rm eff} = \frac{1}{3}\arccos{\frac{q_3}{q_2^{3/2}}}
\end{eqnarray}
where $R_0=1.2 A^{1/3}$ fm and $(J^{\prime}2J0|JJ)$  represents a  Clebsch-Gordan
coefficient.

In Fig.~\ref{fig:def-even}, we have depicted the $\beta_{\rm eff}$ and 
$\gamma_{\rm eff}$ values for the even-even Pt and Os nuclei. A 
monotonic decrease of $\beta_{\rm eff}$ is observed in 
Fig.~\ref{fig:def-even}(a) as one approaches the  $N=126$ shell 
closure. This agrees well with  the gradual shift, from  $\beta\approx 
0.20$ to $\beta\approx 0$, in the global minima of the Gogny-D1M energy 
surfaces (see, Fig.~\ref{fig:pes}). On the other hand, the $\gamma_{\rm 
eff}$ value, plotted in Fig.~\ref{fig:def-even}(b), exhibits a faster 
change with $N$, jumping from  $\gamma_{\rm eff}\approx 25^{\circ}$ (at 
$N=110$) to  $\gamma_{\rm eff}\approx 40^{\circ}$ (from $N=110$ onward) 
in Pt isotopes. Furthermore, the rate change is slower in the Os 
isotopes from $N=114$ up to 118. This behavior of $\gamma_{\rm eff}$ 
confirms that the prolate-to-oblate shape transition takes place. It is 
also consistent with the systematics of the Gogny-D1M energy surfaces.

Similar plots of $\beta_{\rm eff}$ and
$\gamma_{\rm eff}$, are shown in Fig.~\ref{fig:def-odd}  for 
several configurations  close to the ground states of odd-mass
nuclei. As can be seen in
Figs.~\ref{fig:def-odd}(a) (odd-$N$ Pt), \ref{fig:def-odd}(c) (odd-$N$ Os) and
\ref{fig:def-odd}(e) (odd-$Z$ Ir),  the deformation $\beta_{\rm eff}$ 
for each state typically shows a smooth behavior
as a function of $N$. This correlates well with 
the results obtained for even-even systems (see, Fig.~\ref{fig:def-even}).
On the other hand, as in the case of even-even systems, a rapid
change of the $\gamma_{\rm eff}$ value from below to above $\gamma_{\rm
eff}\approx 30^{\circ}$ occurs in some of the states and in each of the 
isotopic chains  shown in panels (b), (d) and (f) of
Fig.~\ref{fig:def-odd}, i.e., in going from $N=109$ to 111 in odd-$N$ Pt
(for the ${3/2}^+_1$ and ${5/2}^+_1$ states), in going from 
$N=113$ to 117 in odd-$N$ Os (for the ${3/2}^+_1$ and ${5/2}^+_1$
states) and in going from $N=110$ to 112 in odd-$Z$ Ir (for the ${11/2}^-_1$ state). 
For those nuclei where $\gamma_{\rm eff}$ changes abruptly, the associated
even-even isotopes also show signs of a 
prolate-to-oblate shape transition (see, Fig.~\ref{fig:def-even}(b)).


\section{Summary and concluding remarks\label{sec:summary}}


In this paper, we have studied the prolate-to-oblate shape phase 
transition in neutron-rich odd-mass nuclei with mass $A\approx 190$. 
Spectroscopic properties have been computed  within a recently 
developed method where most of the parameters of the effective 
Hamiltonian of the IBFM are obtained from an EDF. To this end, the 
$(\beta,\gamma)$-deformation energy surfaces for the even-even core 
nuclei $^{186-200}$Pt and $^{186-194}$Os, spherical single-particle 
energies and occupation probabilities for the corresponding odd-mass 
systems, have been computed within a microscopic EDF framework based on  
constrained mean field HFB configurations obtained with the Gogny-D1M 
parametrization. These quantities have then been used to determine the 
IBFM-2 Hamiltonian. The diagonalization of the IBFM-2 Hamiltonian 
allows to study the properties of $^{185-195}$Pt, $^{185-193}$Os and 
$^{185-195}$Ir. A few coupling constants, for the boson-fermion 
interaction, have been specifically fitted to the low-energy excitation 
spectra for each odd-mass nucleus. However,  those parameters turned 
out to be almost constant or exhibit a gradual variation with nucleon 
number.

Our calculations account reasonably well for the spectroscopic 
properties of the studied odd-mass nuclei. In particular, we have 
identified a clear signature of a shape phase transitions by analyzing 
the systematic trend of the several calculated observables for the 
odd-mass nuclei. For instance, the evolution of the low-lying yrast 
states as well as the effective $\gamma$ deformation parameter exhibits 
significant structural changes at some specific neutron numbers. Our 
results point to the robustness of the prolate-to-oblate shape 
transitions in both even-even and odd-mass nuclei in this particular 
mass region. The present study could be extended further to another 
interesting case, such as those odd-mass nuclei in neutron-deficient Pb 
and Hg regions, which are characterized by a spectacular case of shape 
coexistence phenomena. This would require a major extension of the 
present method, and work along this line is in progress.

\acknowledgments
This work was supported in part by the QuantiXLie Centre of Excellence, a project
co-financed by the Croatian Government and European Union through the
European Regional Development Fund - the Competitiveness and Cohesion
Operational Programme (Grant KK.01.1.1.01.0004).
The  work of LMR was 
supported by Spanish Ministry of Economy and Competitiveness (MINECO)
Grants No. FPA2015-65929-P and FIS2015-63770-P.

\bibliography{refs}

\end{document}